\documentclass[fleqn,usenatbib]{mnras}

\usepackage{newtxtext,newtxmath}

\usepackage[T1]{fontenc}

\DeclareRobustCommand{\VAN}[3]{#2}
\let\VANthebibliography\thebibliography
\def\thebibliography{\DeclareRobustCommand{\VAN}[3]{##3}\VANthebibliography}

\usepackage{graphicx}	
\usepackage{amsmath}	
\usepackage{tabularx}
\usepackage{soul}       

\title[Simulations of Globular Clusters]{Simulations of Globular Clusters Within Their Parent Galaxies: Metallicity Spreads and Anomalous Precursor Populations}

\author[McKenzie \& Bekki]{
Madeleine McKenzie,$^{1,2}$\thanks{E-mail: madeleine.mckenzie@anu.edu.au}
Kenji Bekki,$^{1}$
\\
$^{1}$ ICRAR, M468, The University of Western Australia, 35 Stirling Highway, Crawley, WA 6009, Australia\\
$^{2}$Research School of Astronomy and Astrophysics, Australian National University, Canberra, ACT 2611, Australia\\
}

\date{Accepted XXX. Received YYY; in original form ZZZ}

\pubyear{2021}

\begin{document}
\label{firstpage}
\pagerange{\pageref{firstpage}--\pageref{lastpage}}
\maketitle

\begin{abstract}
Recent observations of globular clusters (GCs) suggest that elemental abundance variations may exist between first-generation (1G) stars. We propose that metal abundance (`metallicity') spreads within GC forming giant molecular clouds (GMCs) can influence the iron abundances of future cluster members. To investigate this, we use original hydrodynamical simulations to model GMC formation in a high redshift dwarf galaxy. Our simulations self-consistently model physical processes such as stellar feedback, dust formation and destruction, and molecular gas formation on dust grains, making them well suited to the study of GMC formation. We conclude that iron abundance variations in GMCs are due to the merging of gas clumps and self-enrichment processes. The metallicity dispersions of GC forming clumps is $\sim$0.1 dex, reflecting a growing number of studies that claim a non-zero dispersion within GCs. The galactic gas fraction is a key parameter for the formation of clumps and the metallicity `floor' observed for both Galactic and extra-galactic GCs are associated with the parent galaxy's capacity to form massive GMCs. Finally, we argue that GMCs have the potential to trap surrounding metal-poor galactic disc stars, which we interpret as a precursor population (0G). These low metallicity stars are representative of the [Fe/H] value of the host dwarf and thus the chemistry of this 0G may be a fossilised record of the parent galaxy. These results depend on the initial metallicity and radial gradient of the galaxy, the threshold gas density for star formation and the star formation prescription.

\end{abstract}

\begin{keywords}
hydrodynamics -- ISM: abundances -- galaxies: ISM -- Galaxy: general -- Galaxy: globular clusters: general
\end{keywords}



\section{Introduction}

Half a century of observations have revealed the complexity of globular clusters (GCs). Early photometric and spectroscopic investigations sparked an enquiry into chemical compositions of clusters (e.g. \citealt{Harding_1962}; \citealt{Norris_freeman1979}; \citealt{Norris_etal1981}) and the confirmation of multiple stellar populations (MSPs) in GCs has since been the subject of countless works (for recent reviews, see \citealt{Bastian&lardo2018}; \citealt{Gratton_etal2019} and \citealt{Cassisi_Salaris2020}). The definition proposed by \citealt{Carretta_etal2010} states that stellar aggregates exhibiting the Na-O anti-correlation could be considered a GC. First generation stars (hereafter 1G stars, nomenclature adopted from \citealt{Piotto_etal2015}) posses chemical signatures similar to that of field stars (e.g. \citealt{Martell_2011}). Second generation or 2G stars exhibit chemical compositions produced by CNO cycling and p-capture processes at high temperatures (e.g. \citealt{Gratton_etal2012}; \citealt{Gratton_etal2019}). GCs which are composed of only these two stellar populations are known as Type I GCs \citep{Milone_etal2017MSP}. Only $~$17\% fail to meet this criteria \citep{Marino_etal2019} and are known as Type II GCs with noteworthy examples including $\omega$ Centauri (\citealt{Lee_etal1999}; \citealt{Bedin_etal2004}), M 54 (\citealt{Carretta_etal2010}; \citealt{Layden_etal2000}), NGC 2808 (\citealt{Harris_1974}; \citealt{Marino_etal2014}; \citealt{D'Antona_etal2016}), M 19 (\citealt{Yong_etal2016}) and NGC 1851 \citep{Carretta_etal2011}. Type II clusters have been hypothesised to be the nucleus of former dwarf galaxies (e.g. \citealt{Bekki_Freeman2003}; \citealt{Da_Costa_2015}).

Light element inhomogeneities within GCs have been the focus of several studies (e.g. \citealt{Hesser_Bell_1980}; \citealt{Norris_1981}; \citealt{Kraft_1994}; \citealt{Marino_etal2008}; \citealt{Gerber_2019}; \citealt{Carretta_2019}; \citealt{Nataf_etal2019} and references therein). \cite{Milone_etal2015M2} pioneered the use of a pseudo two-colour diagram, or chromosome map (ChM), which disentangles various populations harboured by GCs. One result to emerge from this research is the apparent chemical inhomogeneity within the 1G population. The 1G ChM sequence in a significant number of clusters appeared to be either elongated or bimodal; inconsistent with the theory of the 1G being a single stellar population (e.g. NGC2808; \citealt{Milone_etal2015}). The two leading causes for this are variations in either helium (He) or iron (Fe) within the population \citep{Marino_etal2019}. Although well studied in the context of MSP formation (e.g. \citealt{D'Antona_etal2002}), He variations are unlikely to be the cause of an elongated 1G given our current understanding of stellar nucleosynthesis \citep{Milone_etal2018}.

Obtaining accurate measurements of the intrinsic Fe spread within GCs is fraught with difficulties. Systematic uncertainties complicate the unification of multiple data sets and uncertainties for individual measurements are at times comparable to, or larger than, the intrinsic dispersions of clusters themselves. \cite{Carretta_etal2009} found that the upper limit of the scatter of iron in a sample of 19 GCs was less than 0.05 dex and concluded that as far as Fe is concerned, most GCs can still be considered mono-metallic. However, recently \cite{Bailin_2019} compounded a series of studies to allow for more consistent comparisons across data sets. Although small, the dispersions of GCs are measurably nonzero with a median of 0.045 dex and intrinsic dispersions of $\sigma_0$ < 0.1 dex. Using spectra from the APOGEE survey, \cite{Meszaros_etal2020} found that the iron spread in most clusters spans from 0.040 dex to 0.129 dex. However, studies into the metallicity of Magellanic Cloud GCs have not found evidence of this abundance spread (e.g. \citealt{Piatti_2018}).

Stars which are more metal poor than the 1G have also been detected within GCs. The multiple Fe values in Terzan 5 has been problematic in describing its evolution (\citealt{Ferraro_etal2009}; \citealt{Origlia_etal2011}) and the discovery of a third, extremely metal-poor component by \cite{Origlia_etal2013} is difficult to explain with current evolutionary scenarios. Furthermore, the recent validation of $\omega$ Centauri's extended metal-poor tail, which appears to contain two populations \citep{Johnson_etal2020}, lacks a solid rational. The metal-rich cluster 47 Tucane also has a mysterious metal-poor population visible only on the sub-giant branch \citep{Milone_etal2012}. Whether these anomalous observations are linked to the intrinsic spread of the 1G is still a matter of debate.

It has been speculated that a universal mechanism for cluster formation exists  (e.g. \citealt{Elmegreen_Efremov_1997}) and \cite{Harris_Pudritz_1994} proposed that GCs originate from dense cores of supergiant molecular clouds in the early protogalactic epoch. Due to its cosmological implications, GC formation in the context of galaxy evolution is a well-investigated topic. The E-MOSAICS project \citep{Pfeffer_etal2017} builds upon the EAGLE \citep{Schaye_etal2015} galaxy formation model and examines the co-evolution of star clusters and their host galaxies in a fully cosmological context. Other simulations such as those presented in \cite{Kravtsov_Gnedin2005}, \cite{Renaud_etal2017}, \cite{Li_etal2018}, \cite{phipps_etal2019} and \cite{Halbesma_etal2020} also investigate the feasibility of creating bound star clusters; however, the analysis of heavy metal spreads within these possible proto-GCs is less common. \cite{Ma_etal2020} performed self-consistent simulations proto-globular cluster formation in cosmological simulations and recovered a metallicity dispersion of cluster members of $~$0.08 dex in [Z/H]. \cite{Bekki_Tsujimoto_2016} discussed the [Fe/H] spread in the context of GC merging, but only for Type II GCs.

The spatial resolution of many cosmological simulations inhibits the investigation of many critical processes in the ISM such as $\rm{H}_2$ formation on dust grains. In our previous work, \cite{McKenzie_Bekki2021} (hereafter Paper I), we focused on the formation of a 2G within a 1G progenitor. The hydrodynamical code did not consider the formation of $\rm{H}_2$ gas on dust grains, formation and evolution of dust and feedback effects from SNe and dust (e.g., photo-electric heating) in a self-consistent manner.

Therefore, we did not investigate the formation of GC hosting GMCs. Since dust physics can influence the formation of massive clumps (that host GCs) in galaxies \citep{Osman_etal2020DUST}, we adopt a similar code which takes many of these processes in the ISM into account to investigate the chemical abundances.

The purpose of this paper is to analyse the metallicity distribution of a 1G in the context of its parent galaxy. Additionally, we investigate the anomalous metal-poor 0G population which we attribute to captured disc stars. This is made possible by the gravitational pull of massive GMCs during the GC's formation process. In Section \ref{sec:simulations}, we discuss the code used in the present study. We discuss our results and their implications in Sections \ref{sec:results} and \ref{sec:discussion}. Our conclusion is presented in Section \ref{sec:conclusion}.

\section{The Model}
\label{sec:simulations}

To investigate the formation of GMCs within a dwarf disc galaxy, we employ an original chemodynamical simulation code based on smoothed-particle hydrodynamical (SPH) methods. As described in \cite{Bekki_2013}, we self consistently implement the evolution of dust, $\rm{H}_2$ formation and star formation. \cite{Bekki_2015_Preliminary} and \cite{Bekki_2015} build upon this framework to implement dust species as `live dust' particles which can grow and be destroyed through physical processes. The growth of dust directly influences the evolution of ISM \citep{Osman_etal2020DUST}, therefore this code is well suited for studying the chemical evolution of GMC formation in our high redshift galaxy. Each of the aforementioned papers give a comprehensive description of the code used in this study, thus we only provide a short summary here for completeness.

\subsection{Dust Evolution and Chemical Enrichment}

\cite{Osman_etal2020DUST} provides a detailed investigation into the treatment of dust in the present code. The simulation consists of a ‘four-component’ model of dark matter, stars, gas, and dust. Dust particles gravitationally interact with the other three components, which allows for a more physical response to the radiation fields of stars. This code accounts for dust destruction due to supernova (SNe), dust formation by AGB stars SNIa, SNII, growth by the accretion of ISM gas-phase metals, polycyclic aromatic hydrocarbon formation, photoelectric heating of dust grains and dust corrected cooling.The implementation of dust is based on models proposed by \cite{Dwek_1998}.

This simulation tracks the chemical abundances and dust properties of H, He, C, N, O, Mg, Ca, Si, S, Fe and Ba, however for this study we focus on Fe. [Fe/H] is used in order to calculate the cooling rate. Chemical enrichment occurs as a result of star formation and metals are ejected by SNIa, SNII and AGB stars. Nucleosynthesis yields of SNe II and Ia are taken from \cite{Tsujimoto_etal_1995}. As we are not investigating 2G formation through AGB ejecta as we did in Paper I, we use the AGB yields from \cite{van_den_Hoek_Groenewegen_1997}. 
These are the same yield tables used for similar galaxy scaled simulations in \cite{Bekki_2013} and the chemical yields are IMF-averaged and are not dependent on individual AGB stars.

The initial gas-phase metallicity of each particle is specified by radius with respect to the centre of the disc. The starting metallicity used for the majority of the study is [Fe/H] = -1.6 and we use a metallicity gradient of -0.01 dex kpc$^{-2}$ which is consistent with estimates from the Large Magellanic Cloud \citep{Feast_etal2010}. 

\subsection{$\rm{H}_2$ Formation and Dissociation}
$\rm{H}_2$ is the most abundant molecule in the ISM and is assumed to form on the surfaces of dust grains (e.g. \citealt{Cazaux_Tielens2004}). We adopt a model for $\rm{H}_2$ formation which is similar to that used in \cite{Pelupessy_etal2006} where the mass fractions of $\rm{H}_2$ to total hydrogen gas ($f_{\rm{H}_2}$) is determined by local far-UV radiation fields and gas densities. However in our study, the time evolution of dust abundances and compositions are explicitly followed and used to estimate $\rm{H}_2$ formation rates. \cite{Goldshmidt_Sternberg1995} and \cite{Draine_2009} describe the formulas used to derive $f_{\rm{H}_2}$ .

\subsection{Star Formation and Feedback}

The current study employs a H dependent star formation recipe. Three conditions must be satisfied for the conversion of a gas particle into a collisionless `new star' particle. Firstly, the local velocity field must be consistent with gravitational collapse (i.e. $div {\bf v}<0$). To mimic the Jeans instability, the local dynamical time-scale is shorter than the sound crossing timescale. Finally, the local gas density ($\rho_{g}$) exceeds a threshold density for star formation ($\rho_{th}$):
\begin{equation}
    \rho_g > \rho_{th}.
\end{equation}

This star formation recipe has an implicit efficiency of 100\% and the temperature of the particle is not taken into account in the conversion from gas to star.
We mainly investigate models with a threshold of $\rho_{th} = 100 \ \rm{atoms \ cm}^{-3}$, however, we also use models with $\rho_{th} \geq 10^3 \ \rm{atoms \ cm}^{-3}$ which is similar to the typical mass density of the core of a GMC (e.g. \citealt{Bergin_Tafalla2007}). To reduce the computational time, gas-to-star conversion is checked at every $0.01t_{unit}$ (equivalent to the maximum time-step width, $\Delta t_{max}$) which corresponds to a physical unit of 1.41 Myrs. The shortest lifetime of stars for our stellar mass range of 0.1-100 $\rm{M}_{\odot}$ is 3 Myr, which is significantly longer than the 1.4 Myr interval used for checking the gas-to-star conversion. If the minimum conversion time scale was longer than 3 Myr, then it would be possible that SNe explosions could lower the mass densities of gas particles (that should have been converted into new stars before SNe) and we could miss subsequent star formation. However with our current prescription, star formation will be picked up by the simulation and decreasing this timescale will not impact our conclusions.

We run additional models which utilise a $\rm{H}_2$ dependent star formation prescription which introduces a star formation probability ($P_{sf}$):

\begin{equation}
    P_{sf} = 1 - \exp(-C_{\rm{eff}}f_{\rm{H}_2}\Delta t_{max} \rho_g^{\alpha_{sf}-1}),
\end{equation}

where $C_{\rm{eff}}$ corresponds to a star formation efficiency in molecular cores and we assume a value of 0.3. $f_{\rm{H}_2}$ is the $\rm{H}_2$ mass fraction of the gas and $\alpha_{sf}$ is the power-law slope of the Kennicutt-Schmidt law (star formation rate (SFR) $\propto \ \rho_g^{\alpha_{sf}}$ ; \citealt{Kennicutt_1998}) for which we adopt a value of $\alpha_{sf} = 1.5$. Earlier chemodynamical simulations also use this $P_{sf}$ (e.g. \citealt{Bekki_Shioya1998}).

In our fiducial model, SN eject a feedback energy ($E_{SN}$) of $10^{51}$ ergs. 90\% of this energy causes an increase of thermal energy and the remaining energy is turned into random motion. We adopt an adiabatic model for both SN Ia and SN II where thermal energy from each SN can remain adiabatic for a timescale of $t_{\rm{adi}}$. There are many simulations and theoretical works on the adiabatic/Sedov phases of SNe (e.g., equation 8.182 in \citealt{Mo_etal2010}) and we used these works to assume that the Sedov phase can last $10^6$ yr for each SNe. This is a reasonable number for the assumed $E_{SN}$ per SN and the star formation threshold gas density. The influence of SNe on ISM should be investigated in detail without using these approximations in future works with enough resolution to investigate individual SN. A canonical Salpeter stellar initial mass function (IMF; \citealt{Salpeter_1955}) with the slope ($\alpha_{IMF}$) of -2.35 and the upper and lower cutoff masses being 0.1 and 100 $\rm{M}_{\odot}$ is used.

\subsection{Gravitational Dynamics and Hydrodynamics}
\label{sec:grav_and_hydro}
Over $10^6$ particles are used to describe our low mass dwarf galaxy. Therefore, we take a direct-summation N-body approach for the gravitational interaction between DM, stars, gas, and dust. We implement SPH techniques presented in \cite{Hernquist_Katz1989} and the simulations are run on GPUs to optimise the speed of gravitational dynamics calculations. As in \cite{Bekki_2013}, different gravitational softening lengths are chosen for the baryonic and dark matter components (i.e. DM; $\epsilon_{DM}$ and stars and gas; $\epsilon_{b}$) and are determined by the initial mean separation of each component. The mass and size resolutions are given in Table \ref{tab:Model_para}. The SPH smoothing length is time dependent and is determined at each time step so that the number of neighbouring particles is roughly 50-80. As discussed in previous works (e.g. \citealt{Bate_Burkert1997} and \citealt{Tamburello_etal2015}), we assume that the minimum gas clump mass that can be reliably captured by the simulations is 2 SPH kernels which we conservatively estimate to be $\approx 4 \times 10^5 \rm{M}_{\odot}$ based on our gas mass resolution.

The ISM composed of gas and dust is modelled as an ideal gas with the ratio of specific heats $\gamma$ = 5/3. Gas and dust are modelled using the same particles and thus are co-moving. The dust properties in each gas particle can evolve with time depending on the metal accretion onto the gas particle and dust destruction and formation within the gas. The gaseous temperature ($T_g$) is initially $10^4$ K for all models. Radiative cooling processes are implemented using the cooling curve by \cite{Rosen_Bregman_1995} for $T_g < 10^4$K and the MAPPING III code for $T_g \ge 10^4$K \citep{Sutherland_Dopita_1993}. Both codes assume equilibrium which we admit is an oversimplification but is neccecary for our galaxy scale simulations.
We use the gas-phase metallicity ([Fe/H]) for each gas particle to estimate the cooling rate. Dust-phase metals locked up into dust grains do not participate in radiative cooling and thus are excluded from the estimation of gas-phase metallicities. The dust–gas cooling rate is implemented using formulas given in \cite{Tielens_2005}.

A focus of our investigation is the iron abundances of cluster forming GMCs. To estimate the gas-phase [Fe/H] we calculate the gas-phase Fe mass ($m_{\rm{Fe, g}}$) for each particle at each time step via: 
\begin{equation}
    m_{\rm{Fe, g}} = m_{\rm{Fe, t}} - m_{\rm{Fe, d}},
\end{equation}

where $m_{\rm{Fe, t}}$ is the total Fe mass and $m_{\rm{Fe, t}}$ is Fe mass locked in dust particles. We use $m_{\rm{Fe, g}}$ as the total Fe mass of a gas particle because gas and dust evolution are separately evaluated at each time step. This allows for a better estimate of the cooling rates for gas particles compared to simulations which do not include proper dust calculations. We use $m_{\rm{Fe, t}}$ for our analysis of the metallicity of GC forming gas clouds.

\subsection{Dwarf Galaxy Model}

As in Paper I, we assume that a GC forming dwarf galaxy consists of a dark matter halo (DM), stellar disc, and gaseous disc.
The initial total masses of DM, stellar disc, and gas disc
are denoted as $M_{\rm dm}$, $M_{\rm s}$, and $M_{\rm g}$, respectively. We investigate low-mass dwarf disc galaxies with a range of baryonic fractions and where $M_{\rm dm} = [1-5] \times 10^{10} {\rm M}_{\odot}$.

We adopt the `NFW' profile for the dark matter halo (\citealt{NFW_1996}) with a central cusp predicted by the Cold Dark Matter (CDM) model:
\begin{equation}
{\rho}(r)=\frac{\rho_{0}}{(r/r_{\rm s})(1+r/r_{\rm s})^2},
\end{equation}
where $r$, $\rho_{0}$, and $r_{\rm s}$ are the distance from the centre
of the cluster, the central density, and the scale-length of the dark halo, respectively. The virial radius ($r_{\rm vir}$), the scale radius ($r_{\rm s}$), and the `$c$' parameter (=$r_{\rm vir}/r_{\rm s}$) are chosen to be consistent with recent cosmological simulations for the adopted $M_{\rm h}$ (\citealt{Neto_etal2007}).

We treat the mass and size of the galactic bulge in a disc galaxy
as free parameters ($M_{\rm b}$ and $R_{\rm b}$, respectively). Radial ($R$) and vertical ($Z$) density profiles of the adopted exponential stellar disc are proportional to $\exp (-R/R_{0}) $ with scale length $R_{0} = 0.2R_{\rm s}$  and to ${\rm sech}^2 (Z/Z_{0})$ with scale length $Z_{0} = 0.04R_{\rm s}$. The gas disc with a size  $R_{\rm g}=R_{\rm s}$
has the radial scale length of $0.2R_{\rm g}$ and vertical scale length of $0.2R_{\rm g}$. The disc of the present model has $R_{\rm s}=17.5$ kpc and the initial radial and azimuthal velocity dispersions are assigned according to the epicyclic theory with Toomre's parameter $Q$ = 1.5. The gas mass fraction ($f_{\rm g}$) is also a free parameter in the present study.

We also run an additional model which introduces a Milky Way (MW) like potential which we describe in Section \ref{sec:stellar_Halo}. This is used to test the long term evolution of the proto-GC and whether it has the potential to capture disc stars from its parent galaxy.

\begin{table}
	\centering
	\caption{Description of the basic parameters used for our dwarf galaxy model}
	\label{tab:Model_para}
	\begin{tabularx}{\columnwidth}{X >{\centering\arraybackslash}X}
		\hline
Physical property                & Parameter value\\
		\hline
Total mass                       & $\rm{M}_h = 3\times 10^{10} \ \rm{M}_{\odot}$ \\
Baryonic mass                       & $\rm{M}_b = 10^{9} \ \rm{M}_{\odot}$ \\
Structure                        & $r_{vir} $= 42 kpc, c = 16 \\
Initial $\rm{H}_2$ fraction      & 0.01             \\
Initial metallicity              & {[}Fe/H{]} = -1.6 dex    \\
Metallicity gradient             & -0.01 dex kpc$^{-2}$     \\
Initial dust/metal ratio         & 0.4                \\
Star formation density threshold & $\rho_{\rm{th}} = 100 \ \rm{atoms \ cm}^{-3}$ \\
IMF                              & Salpeter ($\alpha = 2.35$) \\
Dark matter softening length     &  $\epsilon_{\rm{dm}} = 323$ pc    \\
Baryonic softening length        &  $\epsilon_{\rm{b}} = 18$ pc     \\
Gas mass resolution              & $2.7 \times 10^3 \ \rm{M}_{\odot}$  \\
Disc star mass resolution        & $675 \ \rm{M}_{\odot}$  \\
		\hline
	\end{tabularx}
\end{table}

\section{Results}
\label{sec:results}

\subsection{Fiducial model}

\begin{figure*}
	\includegraphics[width=\textwidth]{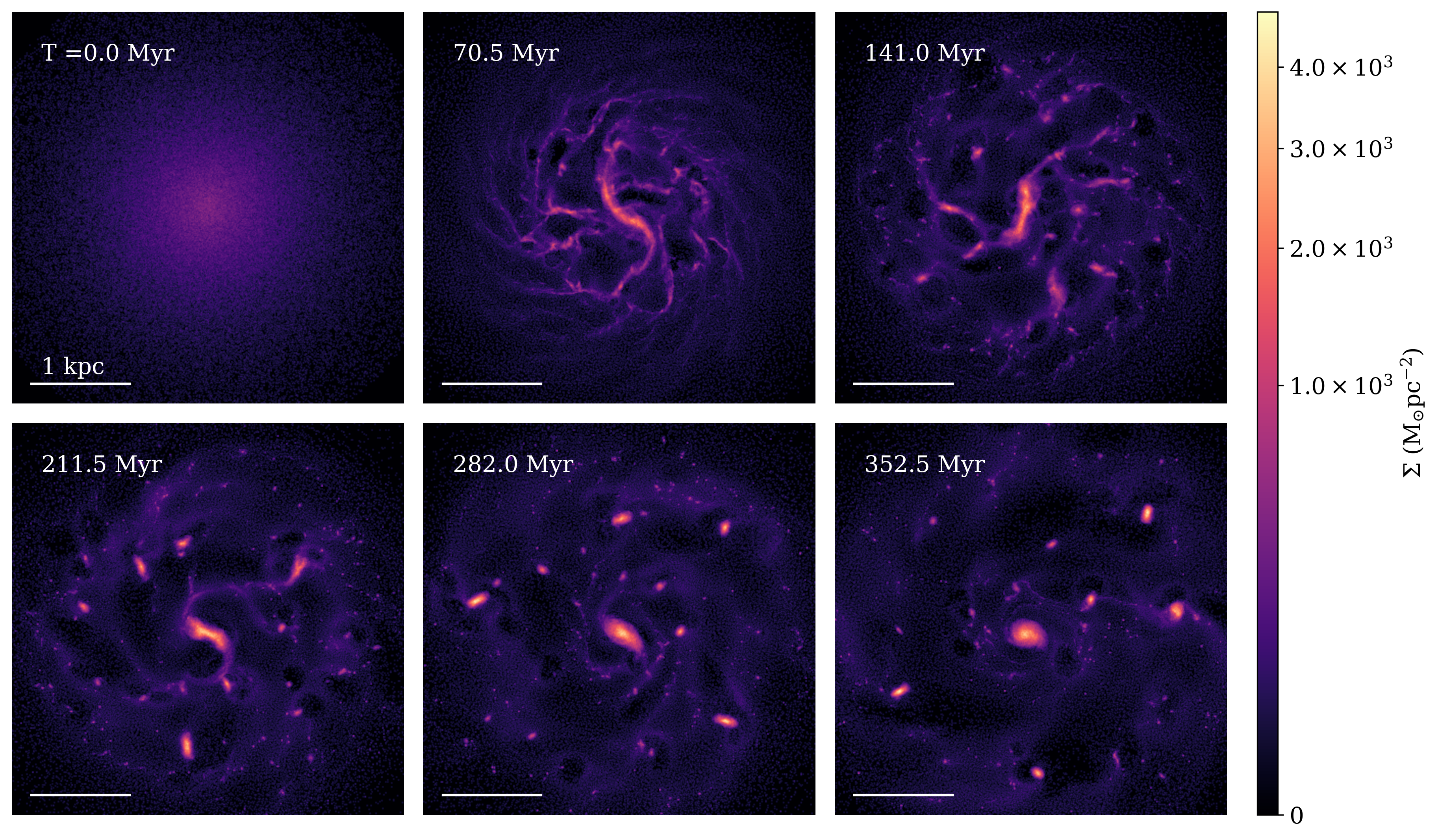}
    \caption{The XY surface density maps showing evolution of gas at six time steps. Each panel encloses a 4 kpc$^2$ region. The 352.5 Myr frame is used for further analysis.}
    \label{fig:gas_evolution}
\end{figure*}

We present the analysis of our fiducial model with parameters summarised in Table \ref{tab:Model_para}. Gas-rich dwarf galaxies with low metallicities ($ \rm{[Fe/H]} < -1.6 $) lead to conditions which promote clump formation. Gas-poor or low baryonic fraction dwarfs result in filamentary structures which do not fragment into clumps. Additionally, galaxies with higher metallicities form fewer high mass clumps than the fiducial model. The galaxy mass-metallicity relation dictates that our large, fiducial galaxy will be unable to explain ultra metal-poor GCs observed in the galaxy (e.g. \citealt{Simpson_2018}). Thus we focus on reproducing a metal rich population of GCs in the present investigation.

Clumps are identified by iterative searching for high density regions in the gas component on a 2D grid. 36 models which ran for a duration of $\sim140$ Myr were used to refine parameters which produced clumps. A further 20 models ran for $\sim350$ Myr to assess clump survival and self-enrichment. Finally, we place our fiducial model into a MW like potential for 1 Gyr to test the long term evolution. We note that our fiducial model from Paper I which used a less massive galaxy with similar radii, would still be conducive to the 1G's formation.

\subsubsection{The Gas Component}
\begin{figure}
	\includegraphics[width=\columnwidth]{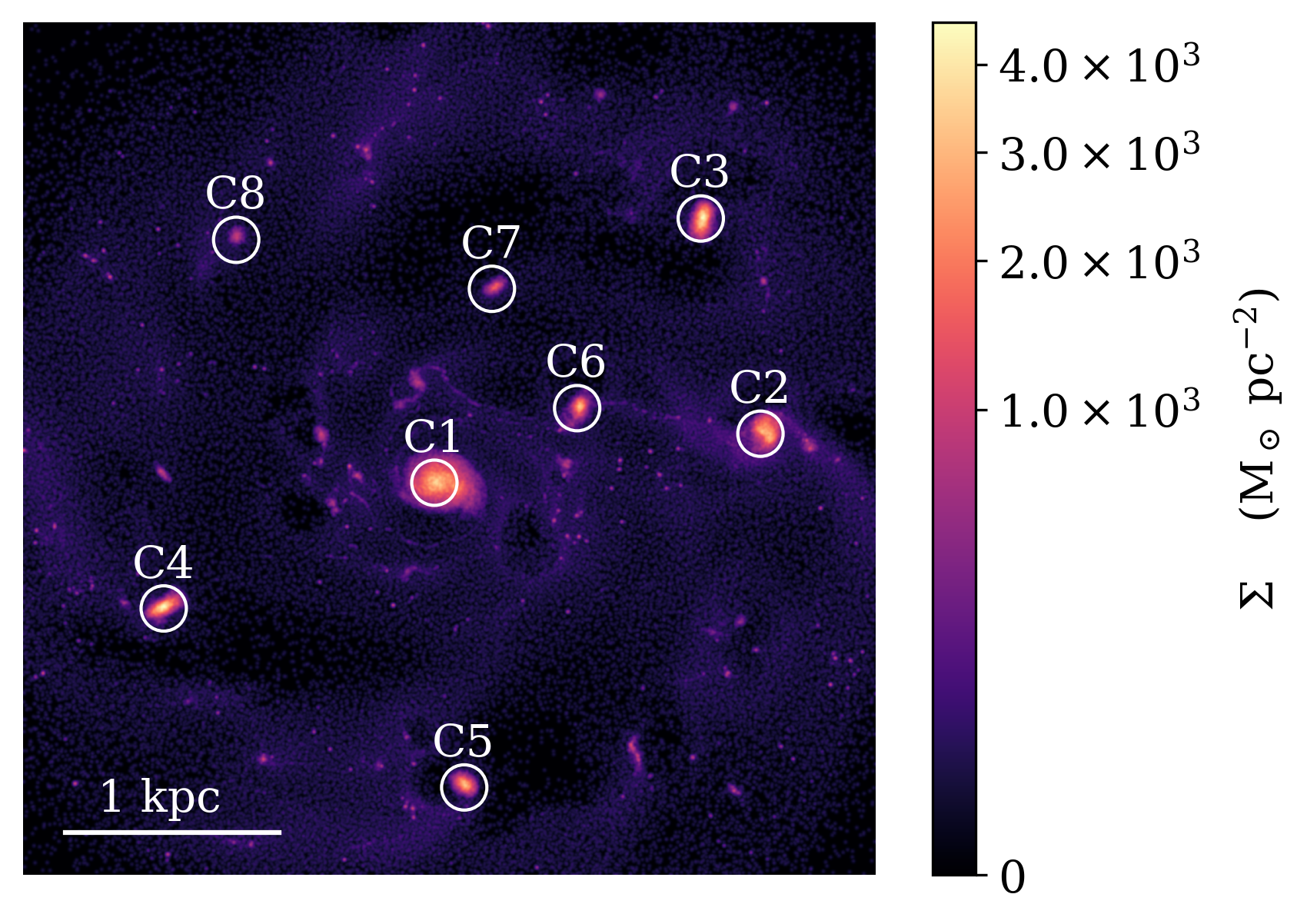}
    \caption{The eight most massive clumps identified for the 352.5 Myr time step from Fig. \ref{fig:gas_evolution}. Each clump is labelled from 1 to 8 and parameters of each clump are listed in Table \ref{tab:fiducial_clumps}.}
    \label{fig:clump_ID}
\end{figure}

Fig. \ref{fig:gas_evolution} illustrates the evolution of the galaxy's gas component. The galaxy starts as an exponential disc rotating in a counter-clockwise direction about the z-axis. After 70 Myr, the gas component exhibits a predominantly filamentary structure which later fragments into clump progenitors. Circular gas voids due to SNe explosions become more pronounced around newly formed clumps after 140 Myrs. Clumps form hierarchically, and the remainder of our analysis focuses on the eight most massive clumps found after 350.5 Myr which we label in Fig. \ref{fig:clump_ID}. There is no strong physical reasons for setting the duration of the simulation to be $\approx$350 Myr. We believe it is long enough to understand how galactic dynamics can influence GMC formation while minimising computational time. This final $\approx$350 Myr snapshot is used for the remainder of our analysis and we list the parameters of each clump in Table \ref{tab:fiducial_clumps}. 

To identify these clumps, we project the particles on a 2D mesh in the X-Y plane with each bin weighted by mass. The mesh spans a region of 4 kpc$^2$ and contains $2.6\times10^5$ bins with a size of 10 by 10 pc. Next, we iteratively select the highest value bins which acts as an initial guess at the locations of GMCs. We set Z = 0 for this initial guess. Once selected, a 3D Gaussian kernel density estimation around our guess of X,Y and Z is used to find the centre of the clump and we calculate the radial density distribution from this central point. To be considered a clump in our fiducial model, the following criteria must be fulfilled:
\begin{enumerate}
\item The radius at which the density drops to $ < 1\rm{M}_{\odot}pc^{-3}$ must enclose a gas mass greater than $5\times10^5\rm{M}_{\odot}$,
\item The clump must have a monotonically decreasing 3D radial density distribution within this radius,
\item The central bin must have a density greater than 10 $\rm{M}_{\odot}pc^{-3}$.
\end{enumerate}

The condition that the radial density distribution must be monotonically decreasing is to avoid any overlap between neighbouring clumps. We set the density cut off value ($1\rm{M}_{\odot}pc^{-3}$) to be in agreement with visual identification of clumps in the gas surface density maps. As the goal of this paper is to analyse the metallicity and velocity dispersion of GMCs which will likely result in dense star clusters, we focus on only high mass clumps. A minimum mass of $5\times10^5\rm{M}_{\odot}$ is required to account for inefficient star formation and mass loss during the 1G formation process. After identification, the clump is removed from the 2D density map to avoid re-selection, regardless of whether it meets the criteria. Finally, the clumps are sorted in descending order based on gas mass, and the top eight are labelled from C1 to C8. This method acts as a preliminary estimate of the number of clumps in the galaxy, and further analysis is performed on the 22 identified clumps to ensure that we have detected a suitable sample.

To study the metallicity and velocity dispersion, we focus on the inner regions of the selected clumps which will likely transform into new star particles. The radius at which the surface density distribution becomes $ 10 \ \rm{M}_{\odot}pc^{-3}$ is used to define the radius of each clump ($R_{c}$). $R_{c}$ is likely an underestimate of the total mass of the clump as we are primarily interested in the metal abundances of the central region of the cloud which will likely be transformed into stars. Despite being a lower estimate of the mass of the clump, C8, our smallest clump is almost 5 times as large as our minimum cloud mass resolvable by our simulations (based on our SPH smoothing length and gas mass resolution discussed in Section \ref{sec:grav_and_hydro}). The mass of the gas component within this region is given in Table \ref{tab:fiducial_clumps}. The four largest clumps have a mass greater than $10^7 \rm{M}_{\odot}$ which validates assumptions made in Paper I that GMCs of this size can form in a high redshift galaxy. GMCs at this scale have also been observed by \cite{Tamburello_etal2015} using the ARGO cosmological hydrodynamical simulations. They found that typical gaseous and stellar masses of clumps ranged from $10^7-10^8 \rm{M}_{\odot}$. However, their investigation used slightly larger galaxies than our fiducial model. \cite{Oklopcic_etal2017} found that in the FIRE simulations, clumps with a mass larger than $10^7 \rm{M}_{\odot}$ accounted for $\approx$20\% of star formation in their galaxies. We note that this study also used far more massive galaxies that our investigation.

The use of $R_{c}$ results in a range of surface densities (assuming $\Sigma = M/\upi \pi R_{c}^2$)) rather than a constant $\Sigma$ assumed in some theoretical models (e.g. \citealt{Bailin_2018}). We find that the mass of the clump scales with radius according to $M \propto R_{c}^{2.8}$.We did not find any correlation between the mass of the clump and its distance from the galactic centre ($R_{gal}$).

\begin{table*}
    \centering
	\caption{The parameters of the eight most massive clumps identified in Fig. \ref{fig:clump_ID}. $\rm{M}_g$, $\rm{M}_d$ and $\rm{M}_n$ are the masses of gas (including $\rm{H}_2$), disc stars and new stars respectively. $\rm{H}_2$ is a subset of the total gas mass. $R_{gal}$ is the distance from the centre of the galaxy. [Fe/H] is the metallicity of the gas and the metallicity dispersion, $\sigma_{[\rm{Fe/H]}}$, is the 1$\sigma$ value for this component. $R_c$ is the radius of the clump which we set as the point where the density reaches 10 $\rm{M}_{\odot}\rm{pc}^{-2}$ and $\rm{n}_g$ is the number of gas particles within this radius. The stellar radius, $R_s$, is taken to be the point where the density is 0.1 $\rm{M}_{\odot}\rm{pc}^{-2}$. $\rm{n}_d$ and $\rm{n}_n$ are the number of particles for the disc and new star components respectively within $R_s$. The clumps without entries had particle counts $<$100 so we exclude them due to lack of resolution. The final column, Offset, gives the distance between the central point of the gas and stellar components for each clump (as discussed in Section} \ref{sec:stellar_comp_offset}).
	\label{tab:fiducial_clumps}
    \begin{tabular}{r c c c c c c c c c c c c c}
    \hline
Clump & $\rm{M}_g$ & $\rm{M}_d$ & $\rm{M}_n$ & $\rm{H}_2$ &  $R_{gal}$ & [Fe/H] & $\sigma_{[\rm{Fe/H]}}$ & $R_{c}$ & $R_{s}$ & n$_g$ & n$_d$ & n$_n$ & Offset\\
    
    ID & ($\rm{M}_{\odot}$)& ($\rm{M}_{\odot}$)& ($\rm{M}_{\odot}$) & ($ \rm{M}_{\odot}$) & (pc) &   &  & (pc) & (pc) & & & & (pc) \\
    \hline
C1         & $3.61\times 10^7$  & $7.66\times 10^6$   & $4.10\times 10^6$  & $1.43\times 10^6$ & 175    & -0.94 & 0.074 & 75 & 150& 13378 & 11349& 2162 & 71  \\
C2         & $1.74\times 10^7$  & $2.78\times 10^5$   & $7.94\times 10^5$  & $7.19\times 10^4$ & 1462   & -1.18 & 0.117 & 59 & 80 & 6433  & 412  & 422 & 100 \\
C3         & $1.53\times 10^7$  & $1.28\times 10^6$   & $5.83\times 10^5$  & $7.28\times 10^4$ & 1600   & -1.13 & 0.092 & 52 & 80 & 5661  & 1904 & 306 & 55  \\
C4         & $1.30\times 10^7$  & $8.67\times 10^5$   & $3.75\times 10^5$  & $1.05\times 10^6$ & 1536   & -1.16 & 0.089 & 49 & 70 & 4816  & 1284 & 193 & 83  \\
C5         & $8.63\times 10^6$  & $3.38\times 10^5$   & $3.44\times 10^5$  & $2.29\times 10^4$ & 1592   & -1.19 & 0.092 & 45 & 70 & 3196  & 501  & 183 & 89  \\
C6         & $6.93\times 10^6$  & $3.31\times 10^5$   & $3.41\times 10^5$  & $3.12\times 10^4$ & 629    & -1.12 & 0.103 & 40 & 60 & 2568  & 265  & 182 & 72  \\
C7         & $1.64\times 10^6$  & -   & -  & 0 & 776    & -1.12    & 0.077  & 26 & - & 607 & - & - & -  \\
C8         & $9.37\times 10^5$  & -   & -  & 0 & 1400   & -1.26    & 0.067  & 24 & - & 347 & - & - & -   \\
    \hline
    \end{tabular}
\end{table*}

\begin{figure}
	\includegraphics[width=\columnwidth]{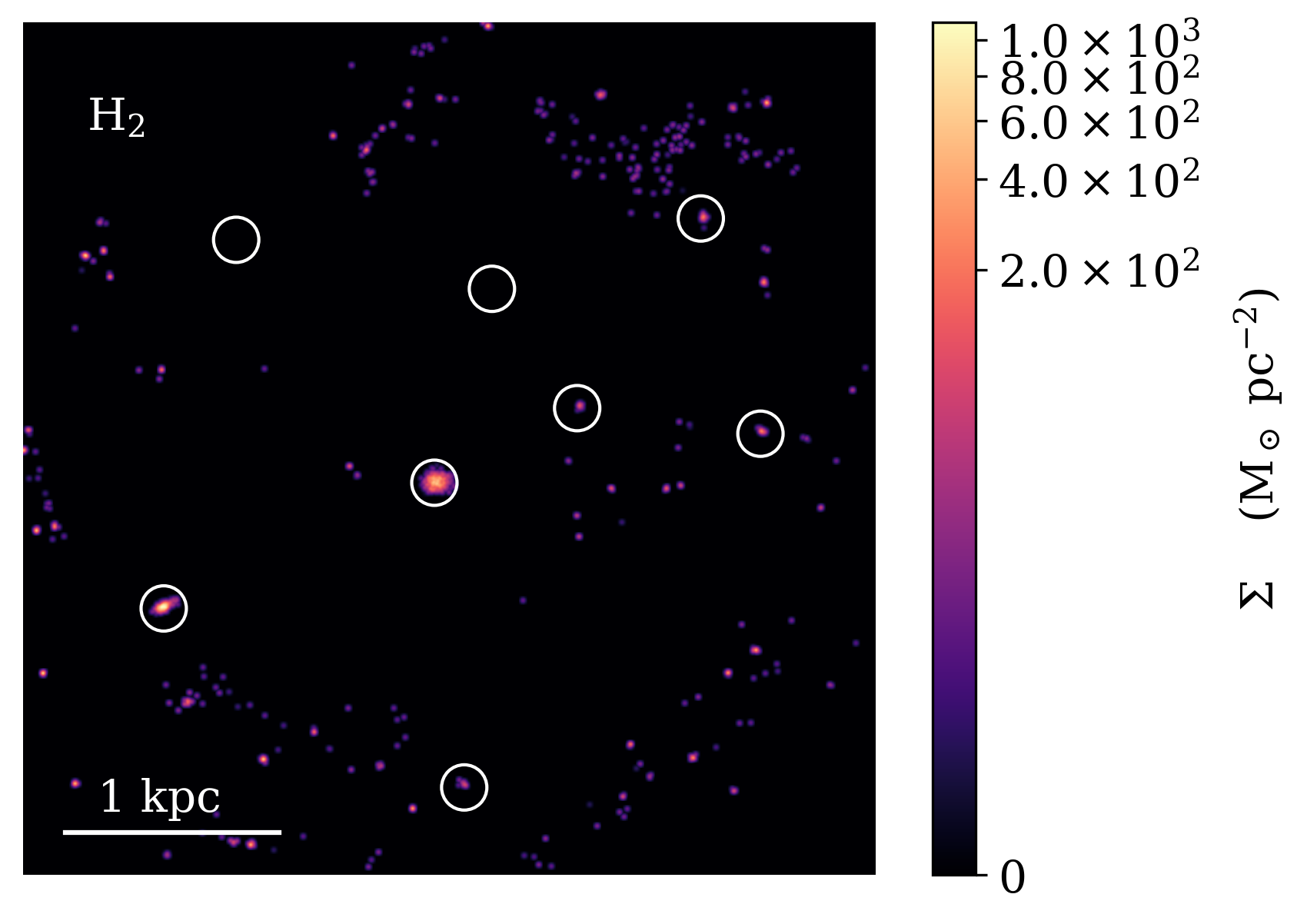}
    \caption{The distribution of molecular hydrogen at the final time step. The white circles are centred on each of the clumps. Excluding Clump 4, the mass of $\rm{H}_2$ correlates with the gas mass.}
    \label{fig:H2_clump}
\end{figure}

The mass of molecular hydrogen ($\rm{H}_2$) within each clump is identified in Fig. \ref{fig:H2_clump}. The $\rm{H}_2$ mass is influenced by several factors including metallicity, dust fraction, gas temperature, the interstellar radiation field and previous star formation consumption. The $\rm{H}_2$ mass correlates with the gas mass in a clump (with the exception of C4). C7 and C8 contained no $\rm{H}_2$ despite having an average gas temperature of $< 150$ K. Small amounts of $\rm{H}_2$ are visible outside the identified clumps suggesting that star formation is still possible in lower density regions. Although many of these regions were analysed during the clump finding stage, they failed to meet the criteria for clump identification.

\begin{figure}
	\includegraphics[width=\columnwidth]{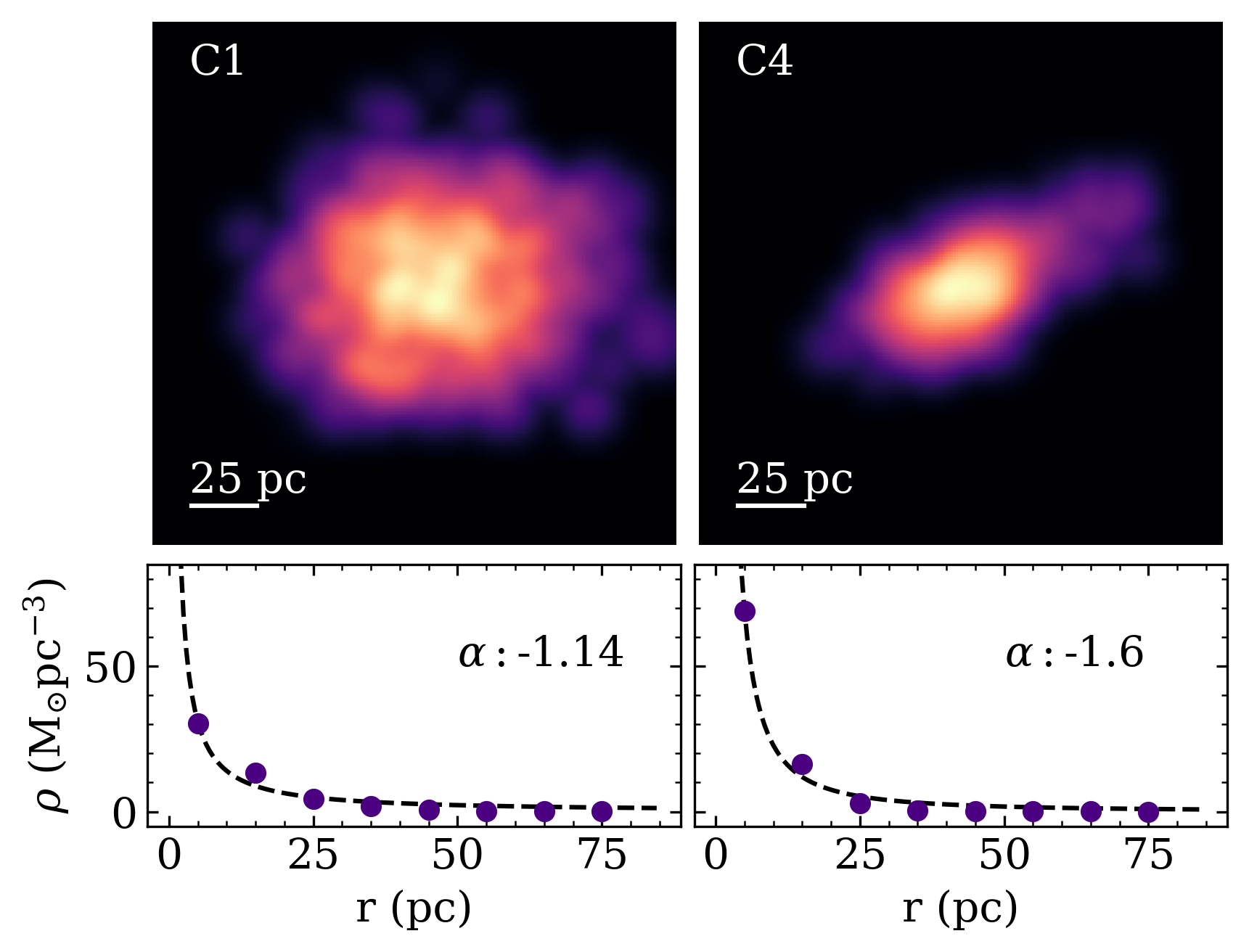}
    \caption{The surface density distributions in the XY direction for the two most massive $\rm{H}_2$ regions and their corresponding 3D radial density distributions.}
    \label{fig:H2_zoom}
\end{figure}

The two clumps with the highest $\rm{H}_2$ mass, C1 and C4, are featured in Fig. \ref{fig:H2_zoom}. The top two panels are the normalised surface density map in the XY direction and the bottom panel shows the corresponding  3D radial density distributions of only $\rm{H}_2$ gas mass. The nuclear clump C1 exhibits a spherical surface density distribution whereas C4 appears more elliptical. Their 3D radial density distributions are reasonably consistent with an isothermal sphere. 

\subsubsection{The Stellar Component}
\label{sec:stellar_comp_offset}

\begin{figure}
	\includegraphics[width=\columnwidth]{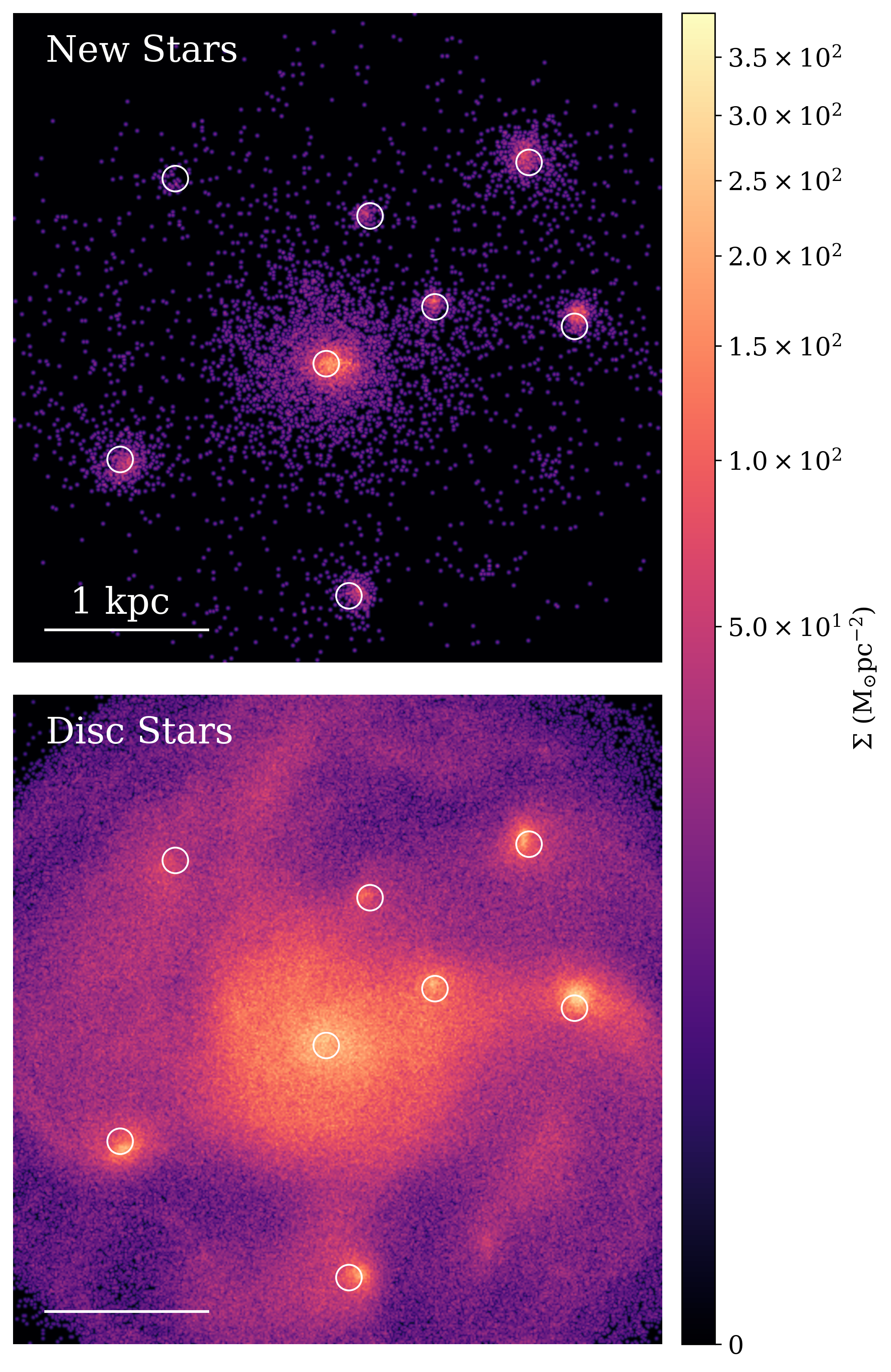}
    \caption{The XY surface density maps of the stellar component in the $\sim 350$ Myr time frame. The white circles represent the location of the 8 largest clumps identified within the simulation. Although there appears to be stellar clusters of stars around these circles, they are not perfectly centred due to ram pressure effects on the gas. The disc stars have a slightly higher density compared to the new stars.}
    \label{fig:new_disc_clump}
\end{figure}

Fig. \ref{fig:new_disc_clump} presents the surface density distributions of the disc stars which were present at the start of the simulation and new stars which formed during a 350 Myr time frame. In both panels, the white circles denote the location of the eight most massive clumps. Stars appear to cluster around the location of these clumps, however, there is an offset between the high density stellar and gaseous regions. We attribute offset to ram pressure between the clumps and the hot gas produced during SNe events. The stellar component is unaffected by ram pressure, and thus the stars appear on the leading edges of the galaxy. On average there is a difference of $\sim 80$ pc between the densest regions of gas and new stars. The individual offsets are given in Table \ref{tab:fiducial_clumps} under the column heading "Offset".

Comparing Fig. \ref{fig:clump_ID} and Fig. \ref{fig:new_disc_clump}, the maximum surface density of the stars is much lower than the gas. Therefore, we define different criteria for the radii of these components. We centre our 3D radial distribution on the densest region of new stars and set the radius to be at the point where $ \rho = 0.1 \ \rm{M}_{\odot}pc^{-3}$. This change is due to the previous limit resulting in a small sample of new stars. The total number of disc and new star particles in C7 and C8 are less than 100, so we choose to exclude them because they can not easily be resolved in our simulation. With this definition, some of the new star masses listed in Table \ref{tab:fiducial_clumps} are large enough to be considered the 1G. However, the radial extent of the new star clumps are much larger than the gas clumps (i.e. comparing the clump radius $\rm{R}_c$ to the stellar radius $\rm{R}_s$) and thus could be influenced by stripping events during the long term evolution of the clusters. The distribution and metallicity of the new stars is heavily dependent on the subgrid physics implementations and our simulation's resolution. Therefore we only focus on the gas in the present investigation. The surface density distribution of the stellar components during the evolution of the simulation are presented in Appendix \ref{sec:spatial_comp_app}.

\subsection{Metallicity}
\label{sec:clump_metal}

\subsubsection{Gas Metallicity and Self Enrichment}

\begin{figure*}
	\includegraphics[width=\textwidth]{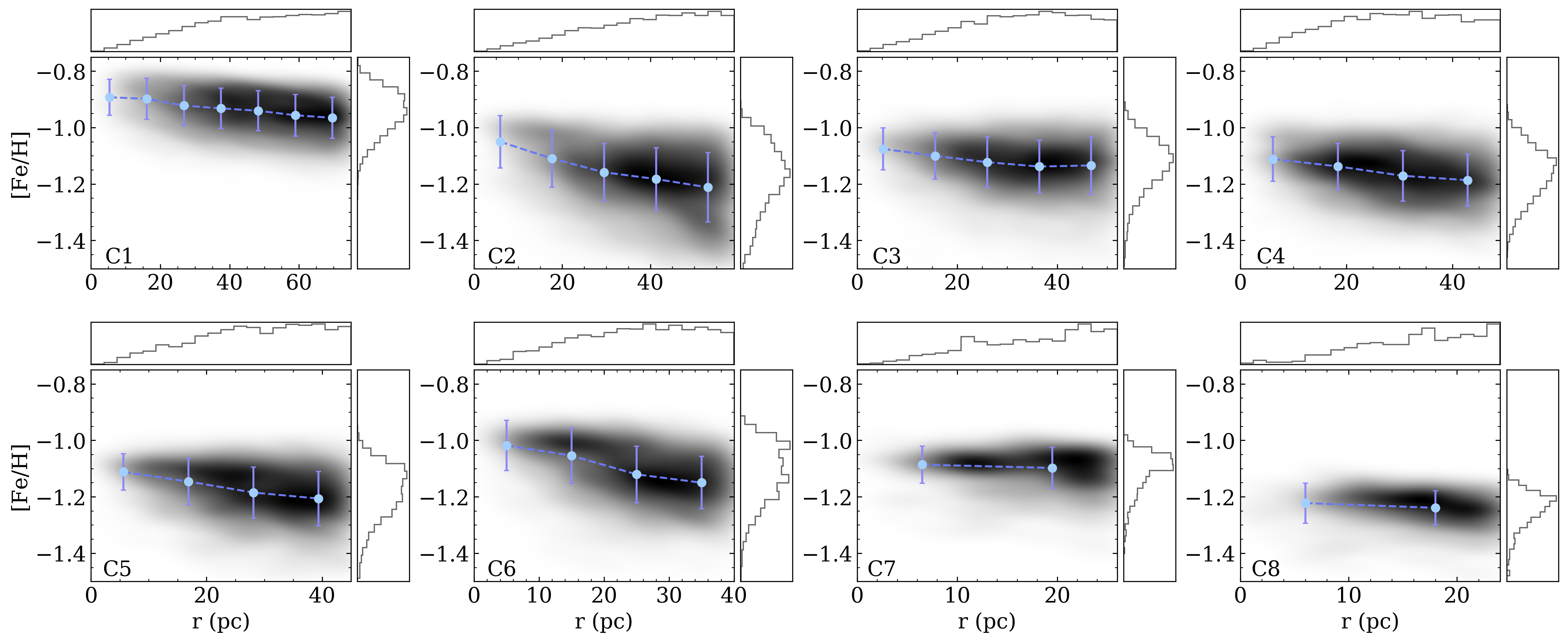}
    \caption{The metallicities of the eight gas clumps identified in Fig. \ref{fig:clump_ID} with labels appearing in the bottom left hand corners of each plot. We consider only the metallicities of particles within the cut off radius. Blue points represent the average [Fe/H] value in 10 pc bins and the error bars show the $1 \sigma$ standard deviation in that bin. Underneath these points, we plot a kernel density estimation to aid in the visualisation of the scatter. The vertical and horizontal axis above each of these plots show the normalised histograms for the radial distribution and [Fe/H] values. Massive clumps are more enriched in iron and a slight metallicity gradient exists within some of the clumps.}
    \label{fig:All_Fe_H_clumps}
\end{figure*}
In this study, we argue that small variations in [Fe/H] are the cause of the extended 1G on the ChMs of GCs. We define [Fe/H] to be the sum of iron in the gas and dust phases. However, the majority of iron is in the gaseous state due to the low dust content of our high redshift galaxy. The radial distribution of [Fe/H] of the GC forming clumps is given in Fig. \ref{fig:All_Fe_H_clumps}. The radius of the clumps is set to be $\rm{R}_c$ defined in Table \ref{tab:fiducial_clumps} and the points and error bars illustrate the mean [Fe/H] and $\sigma_{[Fe/H]}$ in each 10 pc bin. A kernel density estimation of radius as a function of metallicity is shown in grey scale to visualise the dispersion. The top and right histograms represents the normalised radial and [Fe/H] distributions. Metallicities of particles at larger radii typically have a higher degree of scatter and are less enriched than the central bins. 
Variations in helium, another element which could elongate the 1G in a ChM, is on the order of $\sim 0.001$ dex. The very small dispersion agrees with conclusions from observations that He spreads alone can not explain of the ChM spread. This small He variation could be due to the fact that He from AGB stars is spread around 60 surrounding particles and is therefore diluted by the ISM. We also note that we use the IMF averaged AGB models by \cite{van_den_Hoek_Groenewegen_1997} and SNe yields from  \cite{Tsujimoto_etal_1995}. Using different yield tables (for example; \cite{Karakas_2010} for AGB yields or \cite{Sukhbold_etal2016} for SNII ) could increase the amount of He in our GMCs but we don't believe that implementing this change could match observations of the elongated 1G.

In simulations by \cite{Tamburello_etal2015}, they found that the high mass tail of the distribution of GMC masses was mainly the result of GMC mergers, rather than a fragmentation process. In our simulations, we find that the hierarchical nature of GMC formation strongly influences the resulting metallicity, especially for our most massive clumps. C6 is the product of two tidally interacting gas clouds identified during its evolution. An animation of our fiducial galaxy, available as supplementary material, illustrates this merging process. This is evident in the shaded kernel density estimations with a more centrally concentrated metal-rich component. Further merging events may happen in the future evolution of the cluster and may be equivalent to the clumpy accretion phenomenon observed in Paper I. In our current scenario, any high mass gas cloud which experiences several merging events could form a Type II cluster, not just the galaxy's nucleus. C2 also has the largest metallicity gradient spanning 0.5 dex. 

\begin{figure}
	\includegraphics[width=\columnwidth]{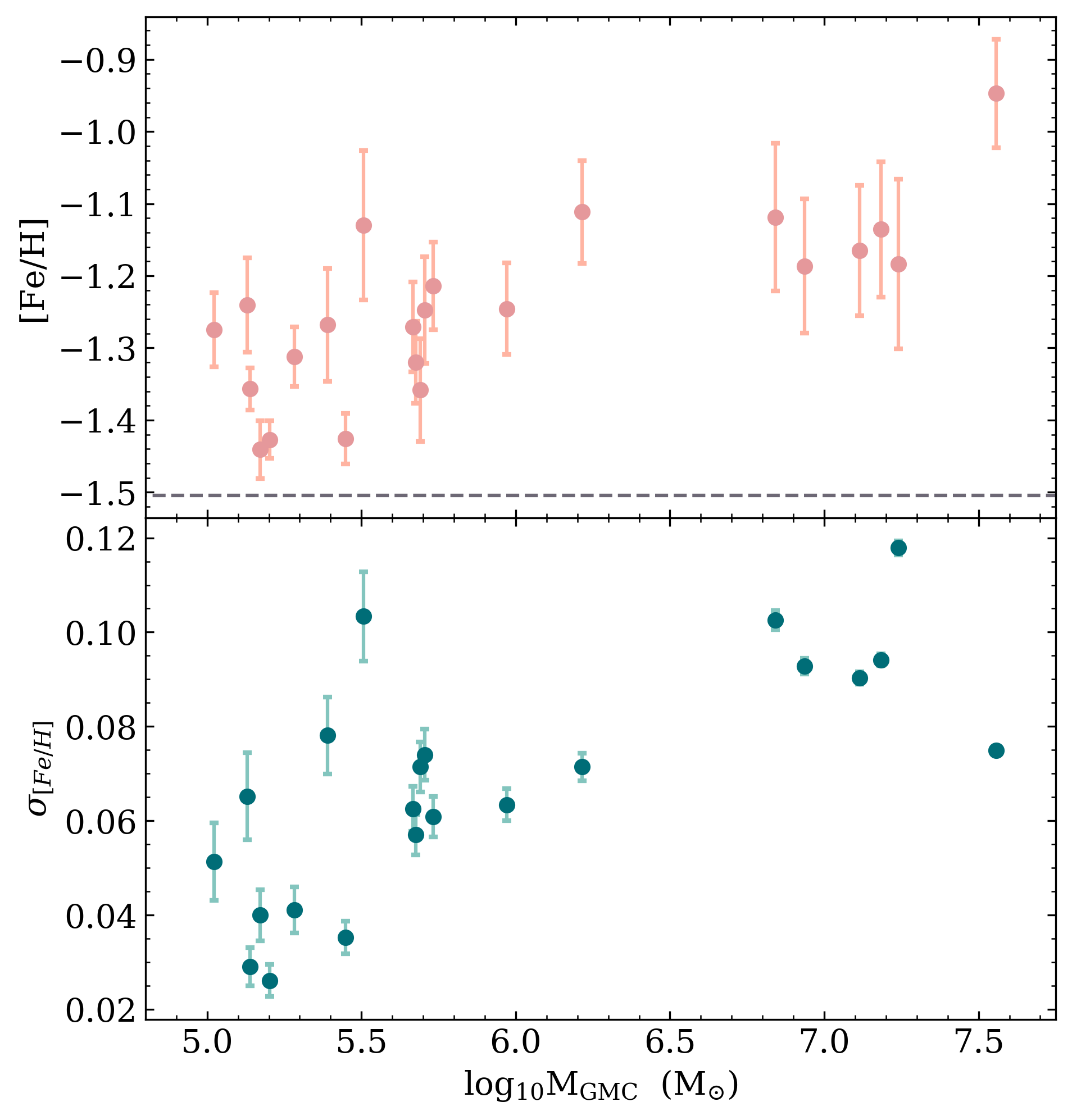}
    \caption{Scaling relation between the mass of the clump and its metallicity for all clumps identified within the galaxy. In the top panel, the error bars denote the $1\sigma$ metallicity dispersion. There is very little variation in the metallicity of the disc star so we show the average [Fe/H] using a grey dashed line. All clumps in the fiducial model show evidence of some degree of self enrichment. The bottom panel shows $\sigma_{\rm{[Fe/H]}}$ with the standard error as error bars. We find that higher mass clumps will have a larger metallicity dispersion.}
    \label{fig:Fe_H_relation}
\end{figure}

There is a correlation between the mass of the clump and its metallicity as shown in Fig. \ref{fig:Fe_H_relation}. This figure plots the mass of our GMC sample against its [Fe/H] value and $\sigma_{\rm{[Fe/H]}}$ in the top and bottom panels respectively. The error bars in the top [Fe/H] panel are the $\sigma_{\rm{[Fe/H]}}$ values and the bottom error bars are the standard error based on the number of particles in each clump. The disc stars surrounding each clump have an average metallicity of [Fe/H] = -1.5 which we show with a horizontal dashed line in the top panel. In both plots, we find a positive correlation between metallicity and gas mass. C1 is significantly more enriched than the galaxy's starting metallicity ([Fe/H] = -1.6). This is evidence of self-enrichment within the clump and we find that the metallicity increases as a function of time. \cite{Bailin_2018} recently discussed formation scenarios of GCs in the context of clumpy self-enrichment. In their scenario, a protocluster cloud fragments into several star-forming clumps. As the stars in these clumps age, they enrich the ISM of other newly forming clusters in the cloud, thus increasing the metallicity and dispersion of the overall cluster. Our galactic scale simulation does not have the resolution to observe this particular phenomenon, however, this should be investigated in future models which can resolve individual star formation on a galactic scale. Their models find a correlation between the mean metallicity of GCs as a function of their stellar mass which is in agreement with observations of metal-poor clusters in the ACS Fornax Cluster Survey \citep{Mieske_etal2010}. Our model also recovers a correlation between these values as seen in Fig. \ref{fig:Fe_H_relation}, however, this involves several assumptions. We do not account for the mass lost due to the formation of new stars and there is a 100\% efficiency in the conversion of gaseous material to stellar material. Importantly, the stellar winds from young stars should not destroy the clump. This allows the new population of stars to deposit enriched material into the gas. Higher-resolution models are required to alleviate these assumptions and properly capture the formation of the 1G in GCs. 

Merging and self-enrichment events both influence the metallicity dispersion of a clump. High mass GMCs are likely to have formed hierarchically and their gravitational potential may allow them to retain enriched material. This is supported by Fig. \ref{fig:Fe_H_relation} which shows that metallciity dispersion correlates with GMC mass. The aforementioned study by \cite{Bailin_2018} and their catalogue of internal metallicity spreads in GCs \citep{Bailin_2019} illustrates that most clusters are not mono-metallic and host internal metallicity spreads of $\sigma_{[Fe/H]} < 0.1$ dex. From our models, we predict that the 1G could have a dispersion as large as 0.12 dex. \cite{Simmerer_etal2013} reported that NGC 3201 had star-to-star abundance variations as high as 0.4 dex, which is too large for our model to explain. However, we do not take possible variations in the metallicity of the 2G into account. Future merging or clumpy accretion events, where smaller star clusters are accreted onto a proto-GC (as discussed in Paper I), could increase the total internal dispersion within the cluster.

\subsubsection{A Higher Star Formation Density Threshold}
\label{sec:High_threshold}

The fiducial model uses a star formation threshold of $\rho_{th} = 100$ atoms cm$^{-3}$. Observations have shown that the cores of GMCs can reach densities as high as $10^5$ atoms cm$^{-3}$ (\citealt{Bergin_etal1996}; \citealt{Bergin_Tafalla2007}), hence we analyse the implications of raising the gas density threshold at which star formation proceeds. Enrichment from the recycling of materials in large star-forming clumps can have a dramatic effect on the resulting metallicity. Massive, long-lived clumps identified in the animation of this model can be $\sim$ 0.5 dex more enriched in iron than the starting metallicity of the galaxy. By restricting the formation of new stars and thus the recycling of metals, the effects of self-enrichment are dramatically suppressed. Using the same parameters as the fiducial model but with a star formation threshold 10 times greater ($\rho_{th} = 10^3$ atoms cm$^{-3}$), we present the equivalent plots for Fig. \ref{fig:clump_ID} and Fig. \ref{fig:All_Fe_H_clumps} for this model in Appendix \ref{sec:rho_th_clumps}. The nuclear cluster in this model (C1B) has a mass of $5.52 \times 10^{7} \ \rm{M}_{\odot}$ and the masses of subsequent clumps range from $4.75 \times 10^{7} \ \rm{M}_{\odot}$ to $4.38 \times 10^{6} \ \rm{M}_{\odot}$ when calculated using the same requirements as the fiducial. The clumps are more massive and contain significantly more $\rm{H}_2$ since less gas has been consumed by star formation. C1B has a metallicity of [Fe/H] = -1.36 dex and $\sigma_{\rm{[Fe/H]}} = 0.08$ dex. The distribution of the clump has a metal-rich tail and spans a range of 0.3 dex; the largest of all models in this study. This is the only clump with evidence of self enrichment yet it is still far less metallic than the nucleus of the fiducial model. All other identified clumps have a metallicity equivalent to the disc stars and a much smaller dispersion. The new stars which formed around the nucleus of the galaxy had a mass of approximately $2.32 \times 10^{5} \ \rm{M}_{\odot}$.

One of the highest precision measurements of iron abundances in a GC come from \cite{Yong_etal2013} in their study of NGC 6752. They found that abundance dispersions were a factor of $\approx$2 times larger than the average measurement uncertainty, thus revealing small but non-negligible metalliciy variations within this Type I cluster \citep{Milone_etal2017MSP}. In their analysis, they found a standard deviation of 0.03 dex; much lower that what can be explained by our fiducial model. However, the high star formation threshold simulation produces several GMC with a metallicity of $\sim$ -1.5 dex and standard deviations $\sim 0.01$ dex (i.e. C2B through to C8B), which can account for observations with such a small iron spread. In this work, the authors suggested that other, if not all, GCs may exhibit comparable abundance variations and correlations to NGC 6752 and that results from photometry exclude He from being the soul contributor to these variations. This aligns with the motivation from our paper that metallicity spreads are due to the chemistry of the GMC imprinting on the GC. As metallicity dispersion correlates with the mass of the cluster and the degree of self-enrichment, this suggests that there should be a mechanism for quenching star formation so that no further self enrichment can occur.

Increasing the threshold density to $\rho_{th} = 10^4$ atoms cm$^{-3}$ completly suppressed star formation and all clumps had a metallicity of [Fe/H] = -1.5 dex and  $\sigma_{\rm{[Fe/H]}}$ on the order of 0.001 dex. Changing the initial [Fe/H] value of the galaxy does not impact the dispersion for these high $\rho_{th}$ models. The SFR for dwarf galaxies at redshifts $z>2$ can range from $1-10 \ \rm{M}_{\odot}/yr$ (\citealt{Thorne_etal2020}). For the fiducial model, the SFR is already below this range at $10^{-1.3} \rm{M}_{\odot}/yr$ and the $\rho_{th} = 10^3$ atoms cm$^{-3}$ model has an even lower SFR of $10^{-2} \rm{M}_{\odot}/yr$. Despite providing interesting predictions, the SFR in this model is unphysically low and thus we keep $\rho_{th} = 100$ atoms cm$^{-3}$ as our fiducial model.

\subsubsection{Alternative Star Formation Prescriptions}
\label{sec:SF_perscriptions}

In addition to testing alternative values of $\rho_{th}$, we also investigate an alternative star formation prescription. This model uses a $\rm{H}_2$ dependent star formation probability ($P_{sf}$) rather than the previous H dependent recipe. As seen in Fig. \ref{fig:H2_clump}, $\rm{H}_2$ makes up a small fraction of the total mass of the GMC. This results in the suppression of star formation and gas recycling.

The accompanying plots for this model are available in Appendix \ref{sec:sf_appendix}. The gas mass is similar to the high $\rho_{th}$ model with the nucleus, C1C, having a mass $5.46 \times 10^{7} \ \rm{M}_{\odot}$ and clumps C2C to C8C ranging from  $1.95 \times 10^{7} \ \rm{M}_{\odot}$ to $2.33 \times 10^{6} \ \rm{M}_{\odot}$. In this case, no clumps show evidence of self enrichment, although the metallicity dispersions for each clump are smaller than the fiducial but larger than the high $\rho_{th}$ model.

\subsubsection{Stellar Metallicity}
\label{sec:Stellar_metal}

The driving mechanism for self-enrichment is the formation and subsequent death of new stars within the cluster. Simulating a 350 Myr time frame allows the galaxy to form many massive clumps, but also is long enough for high mass stars to deposit enriched material into their neighbourhoods. Resolving individual star formation is extremely challenging for a galactic scale simulation. Recently,  \cite{Emerick_etal2019} ran galaxy scale simulations which traced 15 different elements and simulated detailed stellar feedback from individual stars. However, they construct their galaxy model with no initial background stellar population and with a total gas mass of $1.8 \times 10^6 \rm{M}_{\odot}$. We find that their galaxy is smaller than 7 out of our 8 massive clumps analysed in our galaxy model and thus capturing the true formation processes and subsequent metallicity of new stars is beyond the capabilities of our simulation code at this time. We discuss this further in Section \ref{sec:gas_in_cluster}.

\subsection{Kinematics}
\subsubsection{Clump Velocity Dispersion and Internal Kinematics}
\label{sec:clump_dispersion}

We preface this section by stressing the difficulty of analysing GMC scale kinematics within a galaxy scale simulation. The number of gas particles in our smallest clumps limits our ability to accurately determine the rotation and dispersion of the clump. However C8 is still over four times larger than the minimum mass resolvable by our simulations based on our smoothing length estimation. Our softening length ($\epsilon_{\rm{b}}$) is only 1-2 times the value of $\rm{R}_c$, therefore these results should be very carefully interpreted due to the adopted resolution of our fiducial model. As we predict that $\rm{R}_c$ is an underestimation of the total mass of each clump, we include a short description of the rotation and dispersion for clumps in our galaxy, but we do not discuss kinematic radial gradients or structures.

It has become apparent that some GCs show evidence of internal rotation (e.g. \cite{Bianchini_etal2013}; \cite{Kamann_etal2018}; \cite{Milone_etal201847Tuc}; \cite{Bianchini_etal2018}; \cite{Cordoni_etal2020}). In Paper I, we assumed that the 1G population in our model had some net rotation prior to the creation of the 2G. We investigate this assumption by performing a similar analysis to our previous work on our current clump sample. Clumps are divided up into 16 equally sized sectors of a circle in the XY plane, we calculate the average velocity for that segment and then we fit a sinusoidal curve to the values. We extract the amplitudes of the sinusoidal curves for $v_x$, $v_y$ and $v_z$ and summarise these values in Table \ref{tab:kinematics}. The number of gas particles in each clump are given in Table \ref{tab:fiducial_clumps} and were uniformly distributed throughout the 16 sectors. All eight clumps show clear signs of rotation in the X-Y plane with the magnitude correlating with gas mass. Table \ref{tab:fiducial_clumps} also shows that the axis of rotation for all the clumps align with the axis of rotation of the galaxy. Due to the resolution limits of the simulation, there is more uncertainty in the values for the smaller clumps and we do not investigate any kinematic gradients.

\begin{table}
    \centering
	\caption{The values of the x, y and z rotation amplitudes obtained by fitting a sinusoidal curve to the gas component. $\sigma_{g}$, $\sigma_{d}$ and $\sigma_{n}$ are the velocity dispersion for the gas, disc and new star components respectively. }
	\label{tab:kinematics}
    \begin{tabularx}{\columnwidth}{X c c c c c c}
    \hline
     & $v_x$ & $v_y$ & $v_z$ & $\sigma_{g}$ & $\sigma_{d}$ & $\sigma_{n}$\\
     & ($\rm{kms^{-1}}$)& ($\rm{kms^{-1}}$)& ($\rm{kms^{-1}}$) & ($\rm{kms^{-1}}$) & ($\rm{kms^{-1}}$) & ($\rm{kms^{-1}}$)\\
    \hline
C1         & 20.3 & 16.4 & 1.1  & 24.6       & 45.3                 & 31.5 \\
C2         & 16.8 & 16.6 & 1.2  & 17.7       & 23.0                 & 20.0 \\
C3         & 3.9  & 7.4  & 1.1  & 18.1       & 26.0                 & 22.0 \\
C4         & 7.0  & 4.5  & 0.4  & 17.0       & 24.6                 & 20.6 \\
C5         & 5.7  & 5.7  & 0.7  & 13.3       & 17.9                 & 16.4 \\
C6         & 5.2  & 8.6  & 0.8  & 13.0       & 22.8                 & 14.8 \\
C7         & 3.6  & 3.0  & 0.4  & 14.3       & -                    & -    \\
C8         & 2.4  & 3.0  & 1.3  & 5.3        & -                    & -    \\
    \hline
    \end{tabularx}
\end{table}

Only C1 had a sufficient number of new star particles (2162) to repeat the process for the stellar component (see Table \ref{tab:fiducial_clumps}). When centring on the maximum density, the cluster had $v_x$ =  20.6 $\rm{km\ s^{-1}}$, $v_y$ = 15.7 $\rm{km\ s^{-1}}$ and $v_z$ = 3.9 $\rm{km\ s^{-1}}$. These values are very similar to C1 in Table \ref{tab:kinematics} and thus we conclude that the cluster progenitor inherits a similar magnitude of rotation from its parent gas cloud. 

The 1G velocity dispersion $\sigma_{1D}$ (where $\sigma_{1D} = \sqrt{\sigma_{vx}^2 + \sigma_{vy}^2 \sigma_{vz}^2}/\sqrt{3}$) of the clump scales with its mass. Scaling relations by \cite{Larson_1981} found that internal velocity dispersion of molecular clouds is correlated with its size. Although we find that $\sigma_{1D} \propto R_{clump}$, we recover different parameters for this relation. This is due to several factors including the radii of the clumps and that we calculate the dispersion for all gas particles, not just $\rm{H}_2$. The velocity dispersion of the clumps are given in Table \ref{tab:kinematics} and range from $24.6 \rm{kms}^{-1}$ for the nuclear clump to $5.3\rm{kms}^{-1}$ for C8. The resolution of C7 and C8 is too low to determine a reliable $\sigma_{d}$ and $\sigma_{n}$ and so we leave these entries blank.

\cite{vandenBergh_1996} found that merging events are more common in dwarf spheroidal galaxies with a low velocity dispersion. Although the total velocity dispersion for the entire galaxy is quite high ($\sim30 \rm{kms}^{-1}$), the dispersion of the individual clumps is low enough to support merging events as evident from the internal metallicities of some of the clumps. The animation of our fiducial model is available through the supplementary material and can be used to identify merging events. By stepping backwards in time through the animation, one can easily obtain an estimate of the time and location of these events. Clump C2 from the fiducial model merged with another clump 180 Myr after the beginning of the simulation. C6 experienced two merging events; one at 140 Myr and another at 180 Myr. Merging events result in an elongation of the [Fe/H] values as evident in Fig.  \ref{fig:All_Fe_H_clumps} and thus influence the resulting $\sigma_{\rm{[Fe/H]}}$.

\subsubsection{Stellar Captures}
\label{sec:stellar_dispersion}

The dwarf galaxy used in this study starts with a metallicity of [Fe/H] = -1.6 dex. In the fiducial model, self enrichment within high mass gas clumps creates a large metallicity difference between the metallicity of the gas and the stars within the galactic disc. This could explain observations of anomalous metal poor populations of stars existing in GCs which we call the precursor generation, or 0G. This implies that these metal-poor stars represent a fossilised record of the GCs parent galaxy. Terzan 5 is a Galactic GC which is known to have a significant [Fe/H] spread \citep{Ferraro_etal2009}. Recently \cite{Nataf_etal2019} demonstrated that its [Fe/H]=-0.25 population hosts anticorrelations typical of GCs, consistent with the idea that it originated from a normal globular cluster. \cite{McKenzie_Bekki2018} performed hydrodynamical simulations to investigate whether a collision between a GMC and a typical globular cluster could cause the formation of a metal-rich population within a GC. However, this study did not discuss the anomalous precursor metal-poor ([Fe/H] = -0.79) population found by \cite{Origlia_etal2013}. Provided that future simulations can prove that these 0G stars could be retained by the GC during its evolution, we suggest that these metal-poor stars in Terzan 5 are members of this 0G which originate from the parent galaxy.

We make a simple estimate of the fraction of captured disc stars by calculating the escape velocity of the GMC and the absolute velocities of the disc stars once the net rotation of the galaxy has been removed. Because of the high disc star dispersion in C1, only $\approx$20\% of disc stars have velocities less than the escape velocity of the GMC. However for the smaller clumps, 80\% or more of disc stars are not travelling fast enough to escape the clump's gravitational potential. Although future high resolution simulations and more complex calculations are necessary, this supports our hypothesis that these stars have been captured by the GMC.

Given that the disc stars make up a small contribution to the total mass of the system, we predict that observations of a 0G would be rare. Furthermore, observational studies may be biased towards stars which lie closer to the isochrone and not those which show anomalous iron abundances. These simulations suggest that these metal-poor stars are the fossilised records of their parent galaxy. We discuss the implications of capturing disc stars further in Section \ref{sec:stellar_Halo}.

\subsection{Galactic Initial Conditions}
\label{sec:initial_conditions}

\begin{figure}
	\includegraphics[width=\columnwidth]{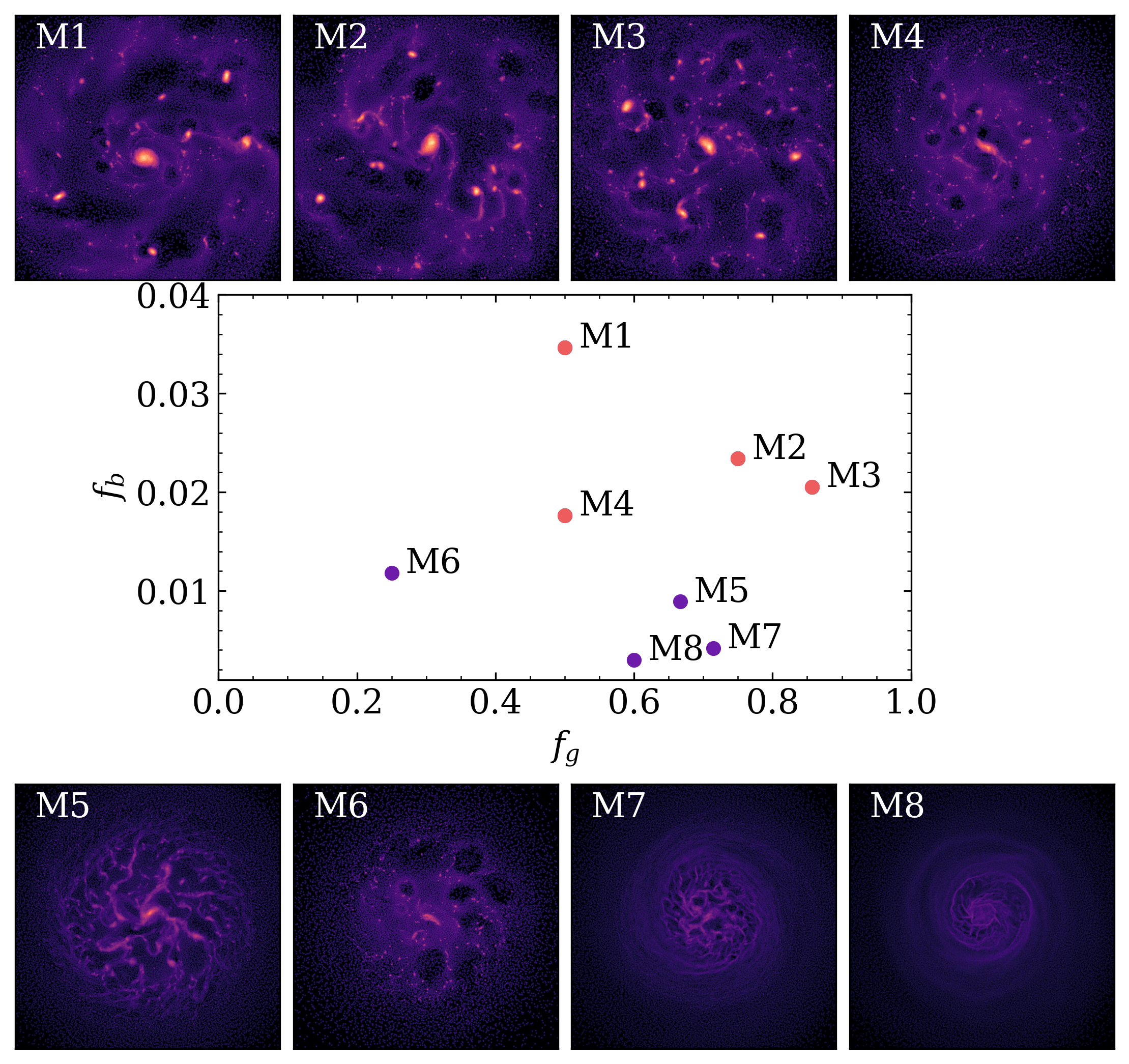}
    \caption{A comparison between the different models analysed during this investigation. The surface density maps surrounding the inner plot illustrates the XY projections of the gas components of the eight models tested. The centre plot shows the gas fraction of the galaxy as a function of its baryonic mass fraction. Orange points represent models which formed gas clumps whereas purple points are for models which did not. The labels in the top left hand corner match up to the labelled points in the inner plot.}
    \label{fig:td7_plot}
\end{figure}

We explore the impact of a galaxy's baryonic and gas mass fractions ($f_b = (\rm{M}_s + \rm{M_g})/(\rm{M_{DM}} + \rm{M}_s + \rm{M}_g$) and $f_g = \rm{M_g}/(\rm{M}_s + \rm{M}_g)$ respectively) on its ability to produce high mass clumps. Fig. \ref{fig:td7_plot} demonstrates the relationship between the total baryonic fraction of the galaxy and the degree of clump formation. Models are labelled from M1 (the fiducial model) to M8 and all have the same dark matter mass of $3\times 10^{10} \ \rm{M}_{\odot}$. Points in orange are models which produced gas clouds large enough to be considered GC progenitors. Models with no evidence of clump formation are plotted in purple. This can be verified by inspecting the XY surface density distributions of each model after 350 Myr surrounding the inner plot. Labels in this plot match up with the labels in the top left hand corners of the surface density maps. Each plot is normalised to the same maximum value to emphasise density variations between the models. M1, M2 and M3 all have the same gas mass of $5.4 \times 10^{8} \ \rm{M}_{\odot}$ (see Table \ref{tab:Model_masses}) and show very defined clumps. Reducing $\rm{M}_g$ results in a decrease in both clump frequency and density as shown by M4, however some of these clumps are still large enough to be considered possible GC progenitors. Variations in stellar mass do not strongly influence the maximum densities of the clumps, but higher fractions of disc stars reduces fragmentation and the number of low mass gas clouds. Over-densities of disc stars occurs for each of the models which show clump formation, regardless of their stellar mass. M5 to M8 all have gas masses $\leq 10^{8} \ \rm{M}_{\odot}$ and show no significant clump formation. The absence of GMC formation may be due to the low availability of gas rather than the gas fraction. Higher mass galaxies may be capable of producing clumps given this combination of $f_g$ and $f_b$, but as our focus is only on dwarf galaxies, we do not test for this. We find that small changes to the gas mass of the galaxy strongly influences clump formation given the current set of parameters in this model. These results have implications for the limit at which a galaxy can form massive GC forming clumps and we discuss this further in Section \ref{sec:Phoenix}.

In Paper 1, we found that the gas fraction of the galaxy could influence the ratios between different populations within a GC; galaxies with a lower gas fraction produced clusters with a lower 2G mass. We wish to correct our conclusion that our result aligned with the work by  \cite{milone_etal2020}. In their study, they did not find a significant relationship between the parent galaxy and the fraction of 1G stars in the cluster. However, our current investigation suggests that only galaxies with a high gas fraction can form a 1G, something we did not explicitly test for in our previous work. This greatly reduces the range of galaxy models we tested and thus we do not expect there to be as much dependence on the parent galaxy's parameters as once thought. Any conclusions made from low mass gas-poor galaxies in our previous study are redundant. However, as the fiducial galaxy from that study is very similar to our current gas-rich models, we believe the scaling relations obtained in Paper 1 are still valid.

\begin{table}
	\centering
	\caption{The masses for the baryonic components for the different models tested. Models each have the same dark matter mass of $3\times10^{10} \rm{M}_{\odot}$ and are ordered by gas mass ($\rm{M}_g$). M1 is our fiducial model. The surface density distributions are shown in Fig. \ref{fig:td7_plot}.}
	\label{tab:Model_masses}
	\begin{tabularx}{\columnwidth}{X X X X}
		\hline
Model ID & $\rm{M}_g \ (\rm{M}_{\odot})$    & $\rm{M}_s \ (\rm{M}_{\odot})$    & $f_b \ (\%)$ \\
        \hline
M1        & 5.39$\times 10^8$ & 5.39$\times 10^8$ & 3.5  \\
M2        & 5.39$\times 10^8$ & 1.80$\times 10^8$ & 2.3  \\
M3        & 5.39$\times 10^8$ & 8.98$\times 10^7$ & 2.1  \\
M4        & 2.69$\times 10^8$ & 2.69$\times 10^8$ & 1.8  \\
M5        & 1.80$\times 10^8$ & 8.98$\times 10^7$ & 0.9  \\
M6        & 8.98$\times 10^7$ & 2.69$\times 10^8$ & 1.2   \\
M7        & 8.98$\times 10^7$ & 3.59$\times 10^7$ & 0.4  \\
M8        & 5.39$\times 10^7$ & 3.59$\times 10^7$ & 0.3  \\
		\hline
	\end{tabularx}
\end{table}

\section{Discussion}
\label{sec:discussion}

\subsection{The Metallicity of the 1G}

The ChM, pioneered in \cite{Milone_etal2015M2} and then extended to a sample of 58 other GCs in \cite{Milone_etal2017MSP}, can help to differentiate different populations of stars in GCs. Each GC's ChM is unique, however work by \cite{Marino_etal2019} removed the metallicity dependence of the cluster to obtain a `Universal ChM of multiple stellar populations in GCs'. During the development of this valuable tool, it was noted that the spread in the 1G was larger than expected from a single stellar population \citep{Milone_etal2017MSP}. To date, there is no phenomenon which has been confirmed to cause this elongation in the 1G. In the Galactic Type II GC NGC 2808, the extended 1G could be caused by a He abundance $\sim$0.03 higher in mass fraction between the populations belonging to the 1G \citep{Milone_etal2015}. Alternatively, one of the populations may instead be enhanced in [Fe/H] and [O/Fe] by $\sim$0.1 dex (\citealt{Milone_etal2015}; \citealt{D'Antona_etal2016}). M3 (another Type II GC; \citealt{Lee_Sneden_2020}) also has an elongated 1G, however \cite{Tailo_etal2019} demonstrated that it cannot be due to He alone. \cite{Milone_etal2018} discussed that if the elongated distribution is due to He, nucleosynthesis processes which alter the He content but not elements involved in standard H burning, such as C and N, would be required. Observations of  C, N, O, Na, Mg, and Al in NGC 2808 by \cite{Cabrera-Ziri_etal2019} confirmed the homogeneous nature of these light elements, adding to the argument that He alone cannot be responsible for this spread. A study by \cite{Martins_etal2020} showed that invoking binaries as the cause of the elongation are unable to reproduce observations. \cite{Marino_etal20193201} echos this result in their analysis of NGC 3201, a cluster with a very large spread in the 1G. Their simulations demonstrated that a large number of binaries would be necessary to account for all the stars in the 1G, thus small variations in metals may also govern its colour spread. However, \citealt{Mucciarelli_etal2015} found that the RGB sample does not show any evidence for intrinsic variations in Fe and conclude that NGC 3201 is a normal, Type I cluster, with no evidence of intrinsic iron spread.

\cite{Marino_etal2019} concluded that primordial iron inhomogeneities at the level of $\sim$0.1 dex could be a possible cause for the spread in the 1G. Our study supports this notion as our eight largest clumps exhibit variations in [Fe/H] large enough to influence the spread of the 1G within the cluster's ChM. More observations and higher resolution simulations are both required to prove this theory, however our theoretical investigations suggest that a primordial spread in iron within the natal GC forming gas cloud is the cause of the extended 1G.

\subsection{Disc Stars Responsible for Stellar Halos and Anomalous Precursor Generations}
\label{sec:stellar_Halo}

Diffuse spherical stellar envelopes have been identified around several clusters (see \citealt{Bekki_Yong2012}; \citealt{Kuzma_etal2016_m2};  \citealt{Kuzma_etal2018}; \citealt{Kundu_etal2020}). A recent study by \cite{Chun_etal2020} used near-infrared APOGEE spectra to search for extratidal stars around M53 and NGC 5053 and discovered that the metallicities of the stars surrounding the GCs were members of the metal-poor population. \cite{Penarrubia_etal2017} suggested that a dark matter halo surrounding a GC could be responsible for creating an extended stellar component. We put forward the suggestion that disc stars captured during the formation of proto-GC gas clouds may be the cause of this phenomenon. Overdensities of disc stars are present in simulations which exhibits clump formation. Provided they survive tidal interactions during the evolution of the cluster, we argue that stars captured from the host galaxy could form a stellar halo with a metallicity lower or comparable to the 1G.

\begin{figure*}
	\includegraphics[width=\textwidth]{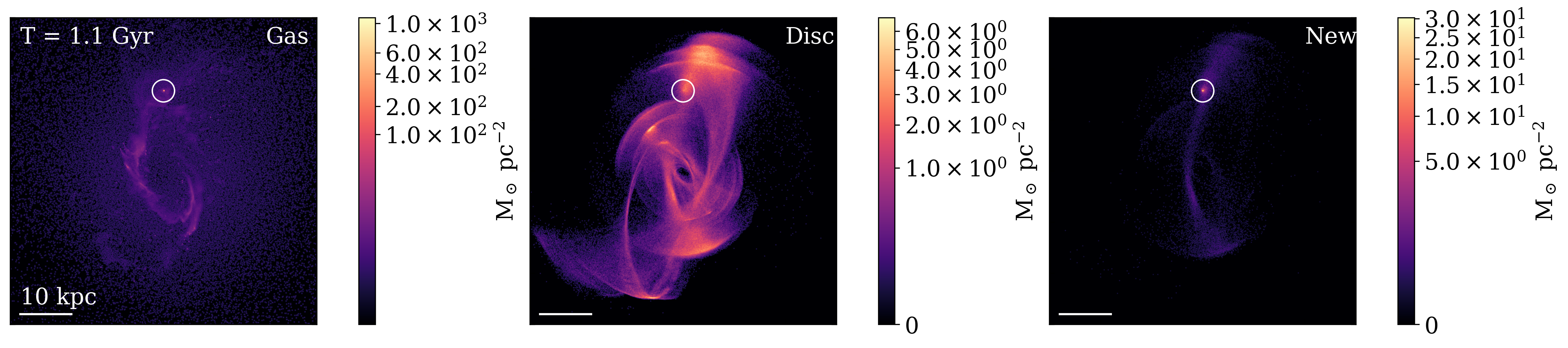}
    \caption{The resulting XY surface density maps after placing the fiducial galaxy in a MW-like potential for $\sim 1$ Gyr. A white circle has been placed around the compact nucleus which survived the destruction of the galaxy. The disc stars have formed a shell like structure and stellar and gaseous streams are visible around the nucleus. Some disc stars surround cluster, however future simulations will be necessary to confirm whether a GC can truly capture disc stars.}
    \label{fig:1gyr_run}
\end{figure*}

In simulations performed in Paper I, we noted that GMCs were capable of capturing surrounding disc stars. The present simulations has shown, for the first time, that GC-hosting GMCs can capture disk field stars from their host dwarf galaxies during their formation. Although this has significant implications on the origin of the anomalous populations of GCs, we can not say with certainty whether captured field stars can be still within GCs after long-term dynamical evolution of GCs. It would be possible that a significant fraction of these field stars (0th generation or 0G) can be tidally stripped by the strong Galactic tidal field during the tidal disruption of their host dwarf galaxies. We intend on studying how GCs with 0G stars dynamically evolve in the Galaxy using self-consistent Nbody simulation codes that can properly model the live Galactic potential (e.g., \citealt{Rossi_etal2016}). However, we briefly discuss this point here using the present simulation code using a fixed Galactic potential.

We mainly investigated the Galaxy mass model (\citealt{Bekki_Tsujimoto_2016}) to discuss the 1 Gyr dynamical evolution of a GC-host dwarf galaxy around the Galaxy. The model parameters for halo, bulge, and disk components of the Galaxy are different between the present and young MW models. The Galaxy in the present MW model is assumed to have a {\it fixed} three-component gravitational potential and a NFW dark matter potential (\citealt{NFW_1996}) with a central cusp as predicted by the Cold Dark Matter (CDM) model. The total mass ($M_{\rm vir}$), the virial radius ($r_{\rm vir}$), the scale radius ($r_{\rm s}$), and the `$c$' parameter (=$r_{\rm vir}/r_{\rm s}$)
for the dark matter halo are set to be $1.3 \times 10^{12} {\rm M}_{\odot}$, 175 kpc, 14.6 kpc, and 12, respectively. The gravitational potential of the Galactic disk is represented by
a \cite{Miyamoto_Nagai1975} potential; 
\begin{equation}
{\Phi}_{\rm disk}=-\frac{GM_{\rm disk}}{\sqrt{R^2 +{(a+\sqrt{z^2+b^2})}^2}},
\end{equation}
where $M_{\rm disk} = 5.0 \times 10^{10} M_{\odot}$, and $a$ = 6.5 kpc, $b$ = 0.26 kpc, and $R=\sqrt{x^2+y^2}$. The following spherical \cite{Hernquist_1990} model is adopted for the potential of the Galactic bulge;
\begin{equation}
{\Phi}_{\rm bulge}=-\frac{GM_{\rm bulge}}{r+c},
\end{equation}
where $M_{\rm bulge}$ = $0.5 \times$ $10^{10}$ $M_{\odot}$, and $c$ = 0.7 kpc. The total masses of the Galactic disk and bulge are significantly smaller than the present-day values, because we consider that GC-hosting dwarfs were destroyed in the early history of the Galaxy formation.

We present only the results of one model in the present study, however, we will conduct a full parameter survey of this investigation in our future papers. The fiducial GC-host dwarf is initially placed at ($x$, $y$, $z$) = (20, 0, 5) (kpc) with respect to the Galactic centre and its initial 3D velocities are $V_{\rm x}$, $V_{\rm y}$, $V_{\rm z}$)= (0, 100, 0) (km s$^{-1}$). In Fig. \ref{fig:1gyr_run}, we show the surface density distributions of the gas, disc and new star components after a period of 1.1 Gyr. With these parameters, the disc stars from the galaxy form a shell-like structure with only the nuclear clump remaining. In Appendix \ref{sec:long_term_ev} we show that after 280 Myr, the galaxy had several more identifiable clumps which later merged into the nucleus in Fig. \ref{fig:1gyr_run}. These clumps merged due to dynamical frictions and/or clump–clump interactions (e.g. \citealt{Noguchi_1999}; \citealt{Inoue_Saitoh2012}; \citealt{Bournaud_etal2014}), however, is possible that more of these GC-hosting clumps may have survived given different initial conditions or a weaker tidal field. After this 1 Gyr period, the dwarf galaxy disc stars form a diffuse halo surrounding the newly formed cluster. Within a 220 pc radius from the centre of the proto-cluster, the disc stars have a mass of $\approx 7 \times10^{4}\ {\rm M}_{\odot}$, however, this number is expected to decrease over the evolution of the cluster. We choose this radius as it includes just over 100 particles after we perform 3$\sigma$ velocity cuts to exclude high velocity particles which are not expected to be bound. We calculate the virial parameters of our system by equating kinetic energy and potential energy (i.e. $\alpha = 2\rm{E}_{kinetic}/\rm{E}_{potential}$) and by using formula described in \cite{Bertoldi_McKee1992}. Using both methods, we recover $\alpha \approx 0.9$. As $\alpha < 1$, this suggested that these stars are indeed gravitationally bound to the cluster.

We hope that future spectroscopic surveys will search for signs of 0Gs within GCs as more evidence of this phenomena would help to verify our GC formation scenario. For example, the metal-poor population in Terzan 5 (\citealt{Origlia_etal2013}) or the metal poor tail of $\omega$ Centauri (\citealt{Johnson_Pilachowski2010}) or M22 (i.e. Fig. 1 from \citealt{DaCosta_Marino2011}) could be potential targets for identifying this population. Additionally, these stars could give us insight into the chemistry of the GCs parent galaxy as the abundance patterns ([Mg/Fe], [Ca/Fe], [Ba/Fe], [Eu/Fe] etc.) of such metal-poor anomalous populations would reflect the chemical evolution of the parent dwarfs. We therefore propose that future observation should investigate the various abundances of potential 0G stars to obtain a better understanding of a GCs connection to its parent galaxy.

\subsection{Gas Remaining in the Cluster}
\label{sec:gas_in_cluster}
It is well established that GCs have a very low neutral gas content (e.g. \citealt{Heiles_Henry1966}; \citealt{Smith_etal1990}; \citealt{Knapp_etal1996}). Evolutionary models predict that stars lose 0.2 $\rm{M}_{\odot}$ during their on the first ascent of the giant branch phase \citep{Tayler_Wood1975}. Mechanisms for expelling gas from clusters include stellar winds (\citealt{Scott_Rose1975}; \citealt{Smith_1996}), accretion by stars \citep{Faulkner_1984}, millisecond pulsars \citep{Spergel_1991}, novae (\citealt{Scott_Durisen1978}; \citealt{Moore_Bildsten2011}) and the sweeping of gaseous medium by the Galactic halo \citep{Tayler_Wood1975}. Additionally, ‘prompt SNIa’ models, where the delay time distribution of SNe Ia is consistent with observational results of SNIa surveys (e.g. \citealt{Mannucci_etal2006}), could be employed as another mechanism for removing such gas. A recent study by \cite{Chantereau_etal2020} on the loss of the intracluster medium in GCs found that inclusion of both ram pressure and ionization is necessary for explaining the very low amount of ionized gas. Furthermore, this rapid clearing of gas is consistent with observations of young massive clusters (e.g. \citealt{Bastian_etal2013}, \citealt{Whitmore_etal2014}, \citealt{CabreraZiri_etal2015}).

In the previous work in this series, we placed a 200 pc HI hole surrounding the 1G of the proto-GC to account for SNe effects. This decision was motivated by observations of large, kpc-scale holes in neighbouring dwarf galaxies \citep{Warren_etal2011}. In the present study, we find that there is a significant gas mass inside and surrounding the clusters after a 350 Myr period. Although multiple SNe events are observed during the simulation (i.e. see the circular gas voids around clumps in Fig. \ref{fig:gas_evolution}) these explosions are not effective in removing the gas from the cluster. 

Star formation within clumps should proceed as detailed in \cite{Bekki_2017}. Bound clusters are formed from molecular clouds with a masses greater than $10^7 \ \rm{M}_{\odot}$ and almost all gas is consumed during cluster formation. Several clumps in our model are above this mass, therefore we predict that given proper modelling of individual stars, 1G formation could consume a significant portion of the remaining gas. Feedback from new stars does not destroy the natal GMC, allowing star formation to precede for much longer time scales. This results in excess self-enrichment within the cluster which poses a problem for our models. Using the moving-mesh hydrodynamics code \textsc{AREPO} \citep{Springel_2010}, \cite{Li_etal2019} discussed how star formation and momentum feedback subgrid models influence the destruction of isolated GMCs due to cluster formation. They found that duration of star formation in simulated GMCs is close to the initial free-fall time of the clouds and that gas expulsion time scale is also dependent on this initial free-fall time. In future simulations, we intend to run models which resolve cluster formation sites on a galactic scale, thus allowing us to include physics such as momentum feedback from individual stars. We hope this enables us to describe how gas expulsion proceeds in GCs formation.

\subsection{Photoelectric Heating and Enhanced Supernova Feedback}

To study the evolution of clumps, we ran variations of the fiducial model which included Photoelectric heating (PEH) calculations and increased supernova feedback. \cite{Osman_etal2020PEH} investigated the effects of PEH of gas in luminous Milky Way type galaxies. In our study, we utilise the same code framework which uses detailed modelling of dust evolution and the time and space varying interstellar radiation field to self-consistently model the effects of PEH. Among several other findings, \cite{Osman_etal2020PEH} concluded that PEH enhances SNe feedback, lessens the abundance of metals and causes the clumps to be less pronounced and have a shorter lifespan over the long term evolution of the galaxy. However, due to the larger dust reservoirs in luminous galaxies compared to our dwarf galaxy models, PEH effects will be less influential in the evolution of our clumps. Models utilising PEH calculations resulted in the suppression of smaller clumps ($<10^5 \  \rm{M}_{\odot}$) but clumps larger than this mass resisted destruction during this 350 Myr time period. Clumps were less enriched compared to the fiducial model, however, there were comparable spreads in [Fe/H]. We only test one set of parameters for PEH and more simulations are required to validate these statements.

After running several higher SNe feedback models, we found that this did not achieve the desired effect of eradicating gas from within the cluster and creating a surrounding HI hole. Violent, high energy explosions and high temperatures prevented the formation of large clumps. Type 1a SNe may be more suitable in removing gas from within the cluster, however as we only simulate a short time frame, these would only become active after the conclusion of the simulation. 

Extensive studies by \cite{Hu_etal2017} and \cite{Hu_2019} involved a detailed analysis of the properties of SN-driven winds and PEH in dwarf galaxies. Their modelling illustrated that the winds of both resolved and unresolved SNe have a dramatic effect on the ISM and the occurrence SN bubbles. Additionally, PEH does not suppress star formation as efficiently as SNe. More comprehensive modelling of SNe in our present simulations will be crucial as it will influence the gas content, enrichment and longevity of any clumps formed.

\subsection{High Redshift Clumpy Galaxies}

Sufficiently cold stellar discs which have a low velocity dispersion are susceptible to gravitational instabilities \citep{Toomre_1964}. Simulations by \cite{Shlosman_Noguchi1993} used three-dimensional collisionless N-body code to model the effect of gas on the global stability of a galactic disk while also discussing the relevance of the Jeans instability. Despite being a small fraction of a galaxy's total mass, they note that the gas can have a notable effect on the global stellar dynamics. Given the increased gas fraction at higher redshift (e.g. \citealt{Daddi_2010}), this effect is expected to become more prominent. \cite{Noguchi_1998} was one of the first to perform numerical simulations of clumpy high redshift galaxies. They found that gas-rich discs of young galaxies becomes gravitationally unstable and fragment into massive, sub-galactic clumps. Our model helps to validate the claim that although gas constitutes a small percentage of the total mass of the cluster, it significantly impacts clump formation. Recent simulations by \cite{Inoue_Yoshida2019} of clumpy galaxy formation in cosmological simulations showed that the clumpiness of galactic discs strongly depends on the equations of state of dense gas. Other factors which contribute to the clumpiness of galaxies have been discussed in detail in the VELA (e.g. \citealt{Moody_etal2014}; \citealt{Mandelker_etal2017}), NIHAO (e.g. \citealt{Buck_etal2017}) and FIRE (e.g. \citealt{Oklopcic_etal2017}) simulations. 

Hubble legacy \citep{Shibuya_2016}, ESO VLT \citep{Genzel_etal2011} and ALMA  \citep{Svoboda_etal2019} observations have provided a wealth of constraints on gas clump mass, size and surface densities for high redshift galaxies. Recent technological advancements have allowed astronomers to probe these galaxies for GC precursors. Gravitational lensing has allowed for the discovery of super-dense star-forming regions with a look back time of over 10 Gyr. Vanzellla et al. (\citealt{Vanzella_etal2017}; \citealt{Vanzella_etal2017_no2}; \citealt{Vanzella_etal2019};\citealt{Vanzella_etal2020}) reported the discovery of several GC precursors including a z = 6.143, $\lesssim 10^6 \rm{M}_{\odot}$ star forming region with an effective radius of less than 13 pc. Additionally, another system at z = 2.37 within the superlensed system dubbed Sunburst arc has a stellar mass of $10^6$ to $ 10^7 \rm{M}_{\odot}$ an effective radius lower than 25 pc (see also \citealt{Chisholm_etal2019}). Given the young age of 2.9 Myr, several clumps in our investigation may fit these criteria. Clumps identified within our simulation are more extended than those within the Sunburst arc, however, they have very similar masses. Thus we support the theory that these star forming regions observed at high redshift could be GC progenitors.

\subsection{GC Metallicity Floor}
\label{sec:Phoenix}

Recently, \cite{Wan_etal2020} reported the discovery of the remains of a tidally disrupted GC belonging to the Phoenix stream. The cluster remnant has an extremely low metallicity of [Fe/H] = -2.7, and \cite{Kruijssen_2020Nature_comment} noted that the galaxy mass-metallicity relationship implies the host of this GC must have been very small. Additionally, \cite{Larsen_etal2020} detected a extra-galactic GC with a metallicity of [Fe/H] =$-2.91\pm 0.04$ dex around our nearest neighbour M 31.

These observations challenge the notion of a metallicity `floor' at [Fe/H]= -2.5 dex for both Galactic and extra-galactic GCs (e.g. \citealt{Beasley_etal2019}). As observations of GCs with [Fe/H] < -2.5 dex are so rare, theoretical studies suggest there should be some minimum host galaxy mass which could support GC formation \citep{Kruijssen_2019}. Fig. \ref{fig:td7_plot} implies that the baryon fraction of the galaxy sets the limit on GMC formation. Our study supports the theory of a metallicity floor as low mass, metal-poor galaxies are unable to form massive, GC forming GMCs.

To investigate this further, we ran an additional model with the initial conditions of M3 (low stellar fraction but high gas fraction; Fig. \ref{fig:td7_plot}) with a metallicity of -3.0 dex. The dwarf has a mass of $3\times10^8  \ \rm{M}_{\odot}$ which we admit is too large for such a low metallicity, however, we assume that the formation of an unusually massive galaxy may be possible. We find that even despite the high gas fraction, this galaxy was incapable of producing clumpy features seen in other, more massive models. The galaxy did produce a nuclear clump with a metallicity of [Fe/H] = -2.66 which is in agreement with the observed remains of the Phoenix stream GC. However, self enrichment from stars in this bulge resulted in a $\sigma_{[Fe/H]} = 0.27$ which is inconsistent with the observed $\sigma_{[Fe/H]}$ of 0.06 dex. This supports theories that Type II GCs with large metallicity spreads originate as the nucleus of such a galaxy. Especially since the metallicity spread in $\omega$ Centauri is said to be 0.205 dex \citep{Meszaros_etal2020}. However, this does not reveal how stars from the Phoenix stream formed.

As our low mass models are unable to form massive GMCs, perhaps a merging event between two low mass high redshift galaxies may be required to achieve the necessary GMC densities to create a GC with such a metal-poor population. We are missing an important piece of the puzzle of how ultra metal-poor GCs can form and future observations from the James Webb Space Telescope will hopefully allow us to answer these questions.

\section{Conclusion}
\label{sec:conclusion}

Our simulations provide physical motivation for metallicity dispersions within the 1G of GCs. We attribute these dispersions to variations in iron abundances within the natal GMCs, which have $\sigma_{\rm{[Fe/H]}}$ on the order of $\sim 0.1$ dex. This value is in agreement with theoretical predictions of the cause of the elongated 1G on the ChMs of GCs \citep{Marino_etal2019}. This metallicity spread is dependent on several factors such as merging events between clumps originating from different locations within the galaxy, and the simulation's star formation prescription. Typical star formation threshold densities used in galaxy scale simulations result in material recycling and self-enrichment of the GMC. This causes a higher [Fe/H] value of the cluster compared to the metallicities of the disc stars. We run preliminary simulations of the destruction of a dwarf galaxy in a MW potential which supports our predictions made in Paper I that GMCs can trap disc stars. We call this population of captured disc stars the 0G precursor population and these disc stars may be the origin of anomalous observations of metal-poor stars in GCs (e.g. \citealt{Origlia_etal2013}). Future surveys should look for these anomalous stars as they may be the fossil records of the chemical evolution of the GCs parent dwarf galaxy.

The gas mass and baryonic fraction of a galaxy is the primary driver of clump formation. Smaller galaxies with low baryonic fractions do not experience high mass clump formation, and from this, we predict that they will not form GC progenitors. This information restricts the number of feasible models used in Paper I and from this study we only trust the results from the gas-rich fiducial galaxy. Given the mass-metallicity relation for galaxies, this observation has direct implications for the formation of metal-poor GCs.

Further investigation into GC formation is required in order to understand their origins. Galactic scale simulations with masses comparable to what was used in this study (i.e. $\sim10^{10} \rm{M}_{\odot}$) which can resolve individual star formation will be the next step in modelling how these clusters formed in the early universe. Nevertheless, these simulations support the long-standing notion that GCs are not mono-metallic and that there is still much to learn about these objects.

\section*{Acknowledgements}

We appreciate the helpful suggestions and thorougher review from our anonymous referee. We are thankful for the time they spent reading our work and discussing our results. MM thanks Dr David Yong for discussing the these results in the context of observations. This study made use of the \textsc{PYTHON} packages \textsc{NUMPY} \citep{van2011numpy}, \textsc{SCIPY} \citep{2020SciPy-NMeth}, \textsc{MATPLOTLIB} \citep{matplotlibHunter:2007}, \textsc{PANDAS} \citep{PANDAS} and \textsc{IPYTHON} \citep{ipythonPER-GRA:2007}. Numerical simulations were run on the Pleiades and \textsc{ozstar} GPU clusters kindly made available to us through the International Center for Radio Astronomy Research (ICRAR) at The University of Western Australia and the Centre for Astrophysics and Supercomputing at the Swinburne University of Technology.

\section*{Data Availability}

The data underlying this article will be shared on reasonable request to the corresponding author.



\bibliographystyle{mnras}
\bibliography{main.bib} 




\appendix

\section{Spatial Distributions for the Remaining Components Within the Simulation}
\label{sec:spatial_comp_app}

\begin{figure*}
	\includegraphics[width=0.9\textwidth]{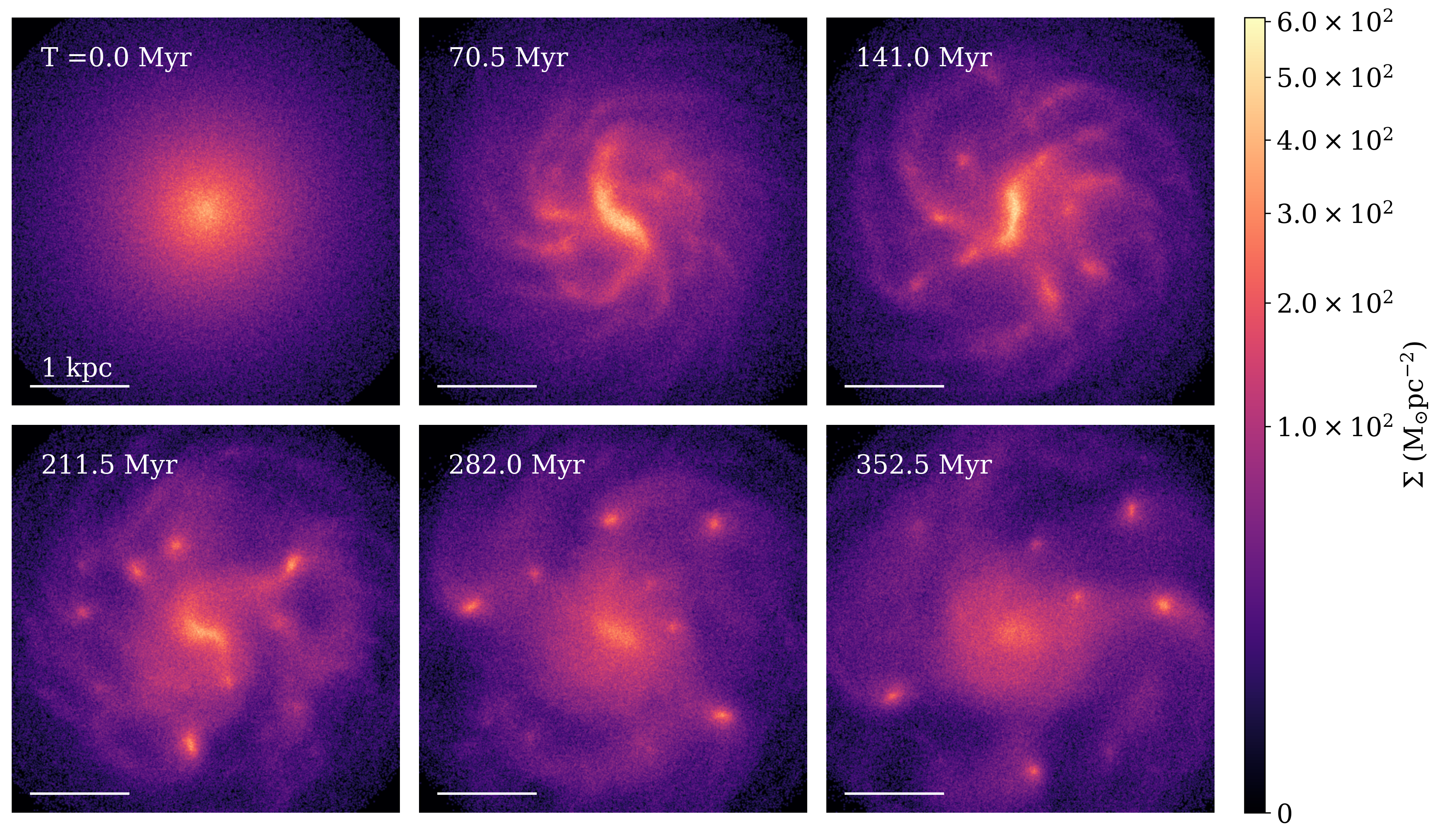}
    \caption{The spatial distribution of disc particles during the simulation. The figure shares the same properties as Fig. \ref{fig:gas_evolution}.
    }
    \label{fig:time_ev_disc}
\end{figure*}

\begin{figure*}
	\includegraphics[width=0.9\textwidth]{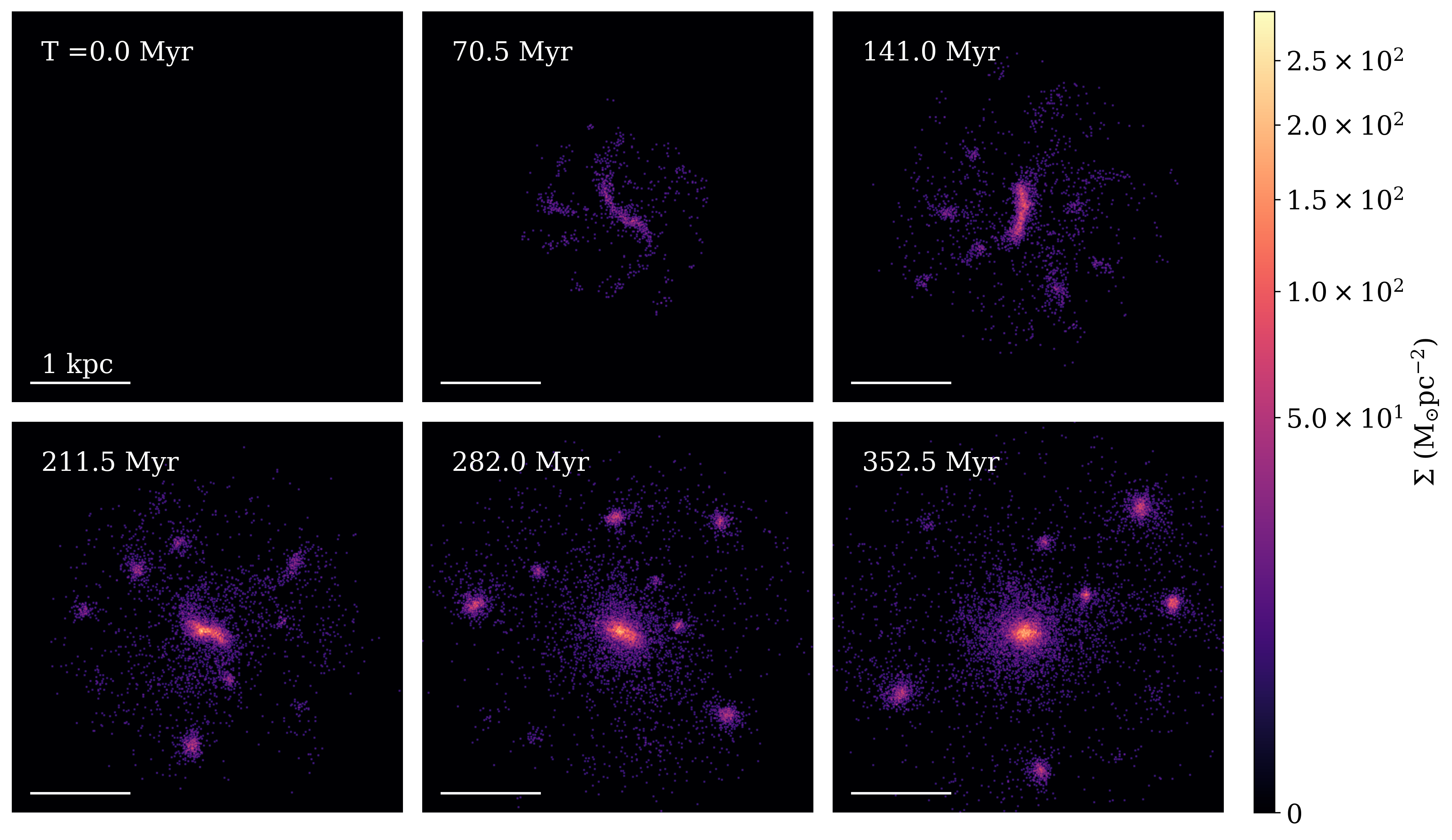}
    \caption{The spatial distribution of new star particles formed during the simulation. The figure shares the same properties as Fig. \ref{fig:gas_evolution}.
    }
    \label{fig:time_ev_new}
\end{figure*}

Figures \ref{fig:time_ev_disc} and \ref{fig:time_ev_new} are the companion figures of Fig. \ref{fig:gas_evolution} for the new stars and disc star components.

\section{High Star Formation Density Threshold}
\label{sec:rho_th_clumps}

\begin{figure}
	\includegraphics[width=\columnwidth]{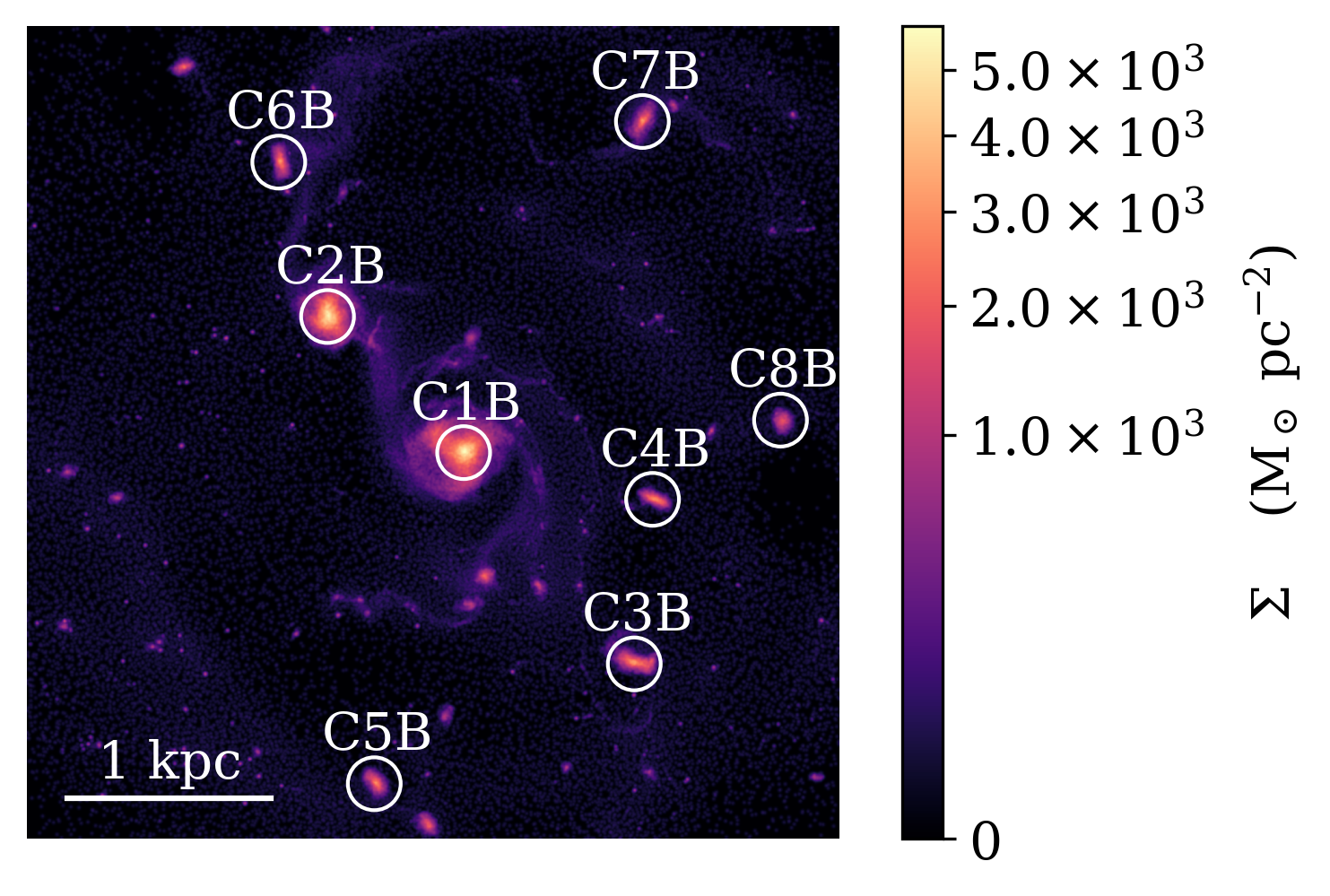}
    \caption{The same as Fig. \ref{fig:clump_ID} but for the high star formation threshold density model. The labelled clumps correspond to the labels in Fig. \ref{fig:All_Fe_H_clumps_high}.
    }
    \label{fig:clump_ID_High}
\end{figure}

\begin{figure*}
	\includegraphics[width=\textwidth]{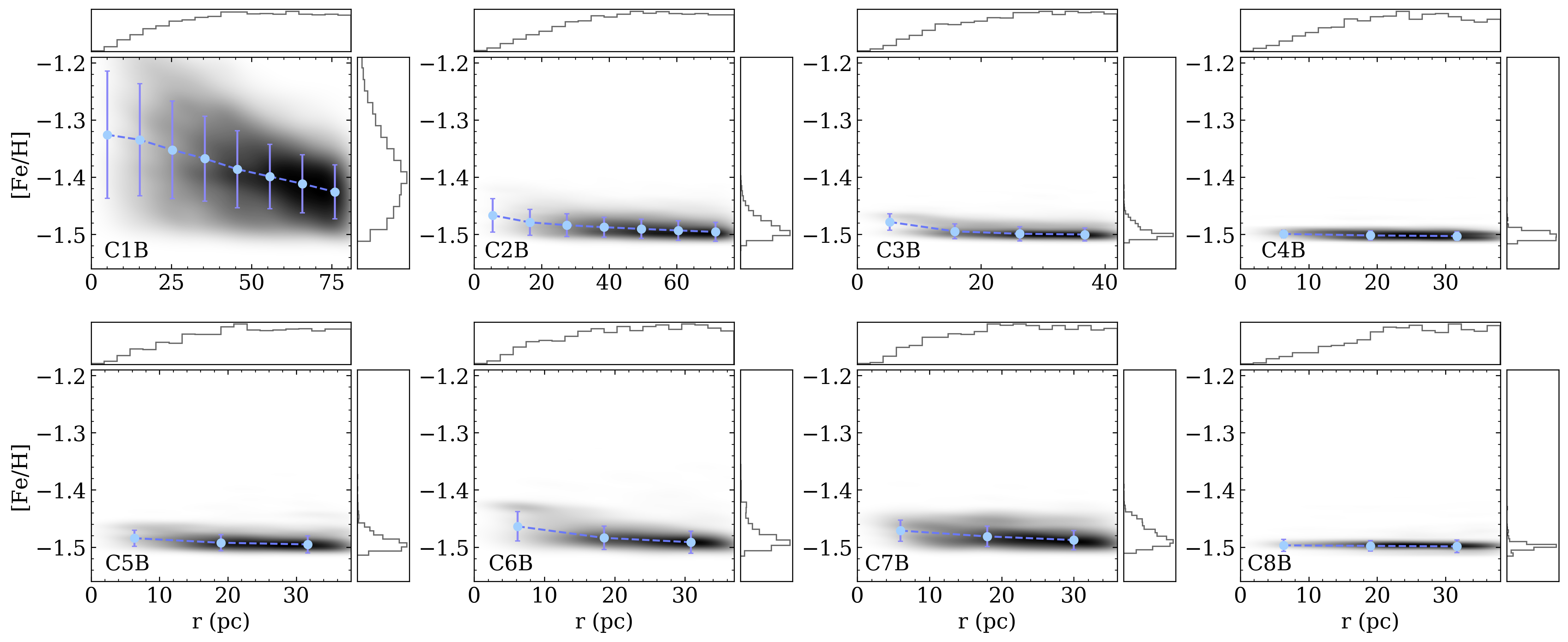}
    \caption{The same as Fig. \ref{fig:All_Fe_H_clumps} but for the eight most massive clumps identified in the high $\rho_{th}$ model. The clumps are less enriched and have a smaller $\sigma_{[Fe/H]}$ compared to the fiducial. The most massive, central clump C1 is consistent with the theory that nuclear dwarfs will have large metallicity variations. All other clumps have metallicities comparable to the disc stars.}
    \label{fig:All_Fe_H_clumps_high}
\end{figure*}

The surface density distribution for the high star formation threshold model discussed in Section \ref{sec:High_threshold}. Clumps are identified in Fig. \ref{fig:clump_ID_High} and the corresponding metallicities are given in Fig. \ref{fig:All_Fe_H_clumps_high}.

\section{Different Star Formation Prescription}
\label{sec:sf_appendix}

\begin{figure}
	\includegraphics[width=\columnwidth]{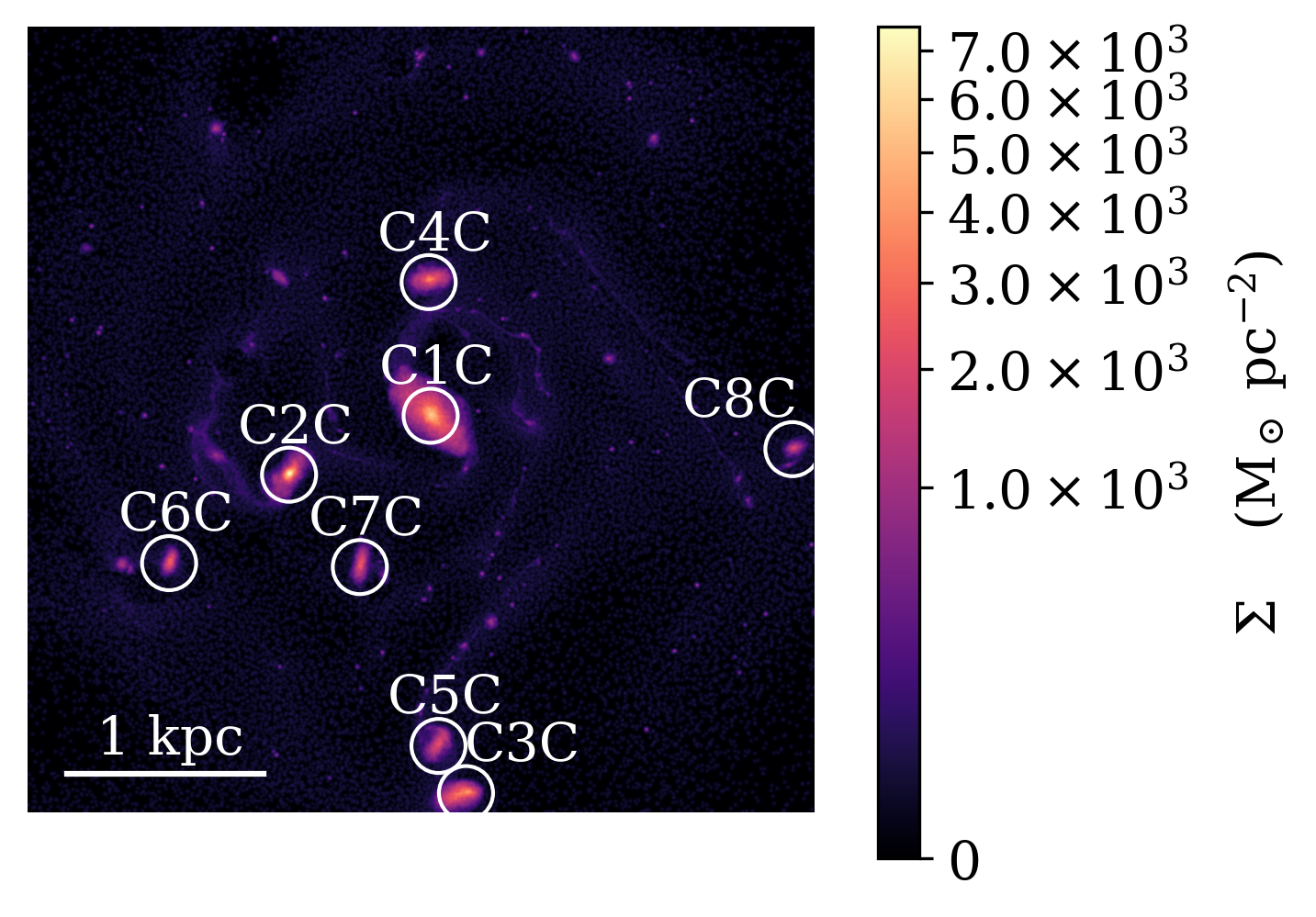}
    \caption{The same as Fig. \ref{fig:clump_ID} but using a different star formation prescription. The labelled clumps correspond to the labels in Fig. \ref{fig:All_Fe_H_clumps_sf}.
    }
    \label{fig:clump_ID_sf}
\end{figure}

\begin{figure*}
	\includegraphics[width=\textwidth]{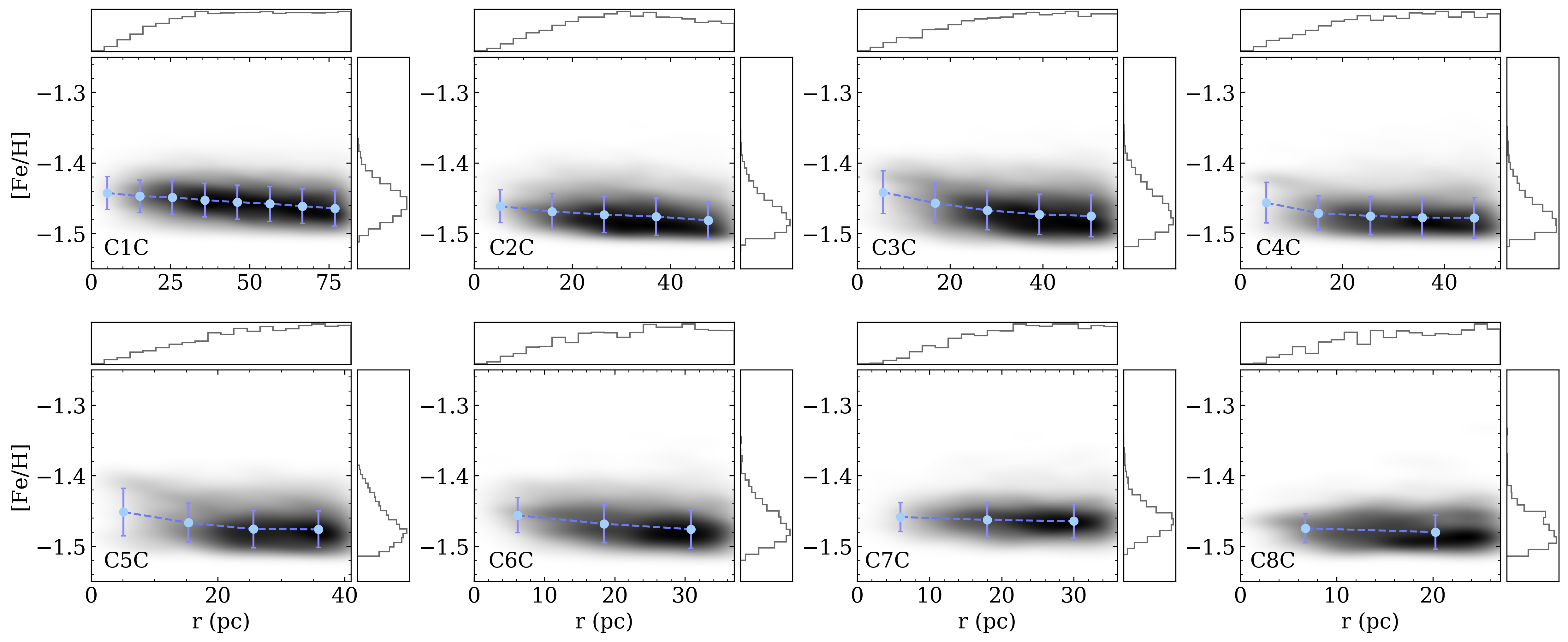}
    \caption{The same as Fig. \ref{fig:All_Fe_H_clumps} but for the eight most massive clumps identified in the different star formation prescription model. The clumps are less enriched and have a smaller $\sigma_{[Fe/H]}$ compared to the fiducial model. All clumps have comparable metallicities to the disc stars and have a smaller metallicity dispersion than the fiducial model.}
    \label{fig:All_Fe_H_clumps_sf}
\end{figure*}

The surface density distribution for the alternative star formation prescription discussed in Section \ref{sec:SF_perscriptions}. As in Appendix \ref{sec:rho_th_clumps}, the labelled clumps are given in Fig. \ref{fig:clump_ID_sf} and metallicities in Fig. \ref{fig:All_Fe_H_clumps_sf}.

\section{Long Term Evolution}
\label{sec:long_term_ev}
\begin{figure*}
	\includegraphics[width=\textwidth]{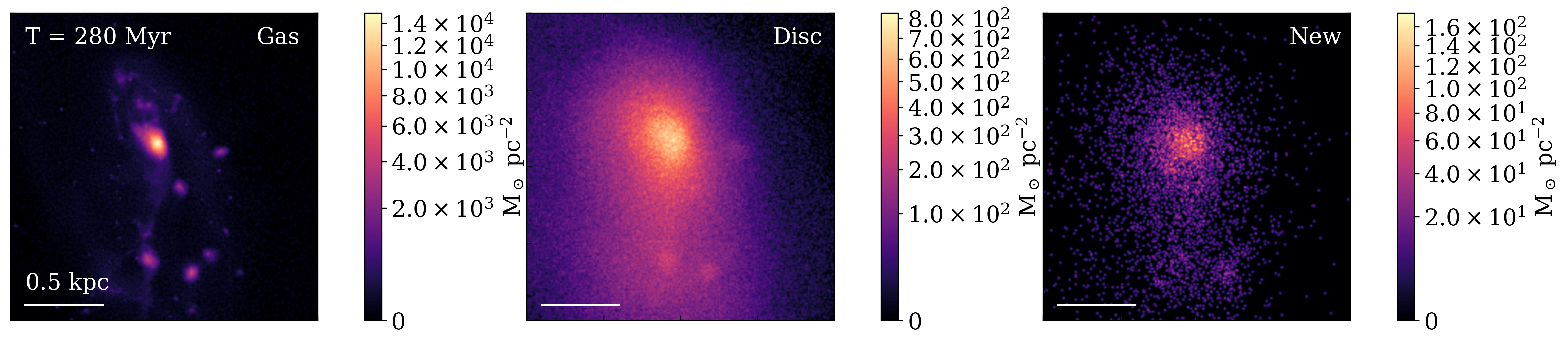}
    \caption{The same as Fig. \ref{fig:1gyr_run} but at T = 280 Myr. The clumps seen in the left most gas panel are later accreted into the nuclear clump.}
    \label{fig:1Gyr_sim_280}
\end{figure*}

The distribution at 280 Myr when the fiducial galaxy is placed in a MW-like potential is shown in Fig. \ref{fig:1Gyr_sim_280}.

\bsp	
\label{lastpage}
\end{document}